\documentclass[twocolumn]{aastex701}

\usepackage{CJK}
\usepackage{tabularx}
\usepackage{booktabs}
\usepackage{inputenc}
\usepackage{textcomp}
\usepackage{subfigure}
\usepackage{amsmath}
\usepackage{multirow}
\usepackage{threeparttable}

\newcommand{\reffig}[1]{Fig.\ref{#1}}

\begin{document}
\begin{CJK*}{UTF8}{gbsn}

\title{Mass stratification in the globular cluster system revealing the assembly history of the nearest S0 galaxy NGC 3115}

\author{Haoran Dou (窦浩然)}
\affiliation{School of Physics and Astronomy, Beijing Normal University Beijing 100875, PR China}
\email{202321160002@mail.bnu.edu.cn}

\author{Hao Li (李昊)}
\affiliation{Department of Astronomy, University of Science and Technology of China, Hefei 230026, PR China}
\affiliation{School of Astronomy and Space Science, University of Science and Technology of China, Hefei 230026, PR China}
\email{lh123@mail.ustc.edu.cn}

\author{Hong-xin Zhang (张红欣)}
\thanks{Corresponding author. E-mail: hzhang18@ustc.edu.cn}
\affiliation{Department of Astronomy, University of Science and Technology of China, Hefei 230026, PR China}
\affiliation{School of Astronomy and Space Science, University of Science and Technology of China, Hefei 230026, PR China}
\email{hzhang18@ustc.edu.cn}

\author{Heng Yu (余恒)}
\thanks{Corresponding author. E-mail: yuheng@bnu.edu.cn}
\affiliation{School of Physics and Astronomy, Beijing Normal University Beijing 100875, PR China}
\email{yuheng@bnu.edu.cn}

\author{Huiyuan Wang (王慧元)}
\affiliation{Department of Astronomy, University of Science and Technology of China, Hefei 230026, PR China}
\affiliation{School of Astronomy and Space Science, University of Science and Technology of China, Hefei 230026, PR China}
\email{whywang@ustc.edu.cn}

\begin{abstract}
    Galaxy formation and evolution is hierarchical. The most massive galaxies are thought to form their central regions early through violent dissipational processes, then grow inside-out by accreting smaller satellites. While widely supported, direct observational confirmation of this process in individual galaxies remains lacking, except for the Milky Way. We present a detailed analysis of globular cluster (GC) candidates within a $70^\prime$ ($\sim190$ kpc) radius around the nearest S0 galaxy, NGC 3115, using images in \textit{g,r,z} bands from the DESI Legacy Imaging Surveys and data from Gaia. We report the discovery of mass stratification in the GC system (GCS), evident in two ways: first, the effective radius of the GCS increases monotonically from the bright to faint end, up to the detection limit near the turnover magnitude of the GC luminosity function (GCLF); second, the GCLF shows fainter turnover magnitudes and smaller standard deviations at larger galactocentric radii. This stratification cannot be readily explained by radial migration or tidal dissolution, but most likely reflects the hierarchical assembly of NGC 3115's stellar halo, with later-accreted satellites deposited across broader galactocentric distances. This interpretation is supported by cosmological simulations of subhalos with comparable mass and bulge-to-total mass ratios and is consistent with the negative color gradients observed in the GCS. Additionally, we identify several substructures within the GCS, indicating ongoing assembly of NGC 3115.  This work highlights the power of GCS as tracers of galaxy assembly and sets the stage for upcoming space-based wide-field imaging surveys to constrain the assembly of massive galaxies.
\end{abstract}


\section{Introduction} \label{sec:intro}

In hierarchical cosmological models, the formation of the most massive galaxies typically follows a two-phase inside-out assembly process: an early dissipative phase that rapidly builds up the central region, followed by the gradual accretion of satellite galaxies that assembles the outskirts \citep{Naab2009,Oser2010,Rodriguez-Gomez2016}. This scenario is broadly consistent with the apparent structural evolution of massive early-type galaxies since redshift $z \sim 2-3$ \citep{vandokkum2010,Buitrago2017}. However, direct and unequivocal observational evidence for this process remains elusive. On one hand, structural evolution inferred from lookback studies may be influenced by the well-known progenitor bias \citep{Cassata2013,Suess2020}, i.e.,
galaxy samples selected at different cosmological redshifts may not originate from the same progenitor population and so their observed difference may not necessarily reflect evolutionary changes. 
On the other hand, while recent archaeological analyses based on spatially resolved spectroscopy of nearby massive galaxies appear to challenge the significance of inside-out mass growth \citep[e.g.,][]{Avila-Reese2023}, accurately reconstructing star formation and chemical evolution histories from integrated spectra remains notoriously difficult. Moreover, spectroscopic observations rarely extend to the faint stellar halo, which is more sensitive to satellite accretion than the galaxy’s central region. 

Globular clusters (GCs) are among the oldest stellar systems in the local universe and thus serve as fossil records of early star formation and evolution of galaxies.
Their high mass and compact structure allow them to survive until the present day, making them powerful tracers of their host galaxy’s assembly history, particularly in the outskirts where diffuse stellar light is challenging to detect \citep{Brodie_2006,Beasley_2020}.
With a stellar mass of $\simeq1.1\times10^{11}\ \rm M_{\odot}$ \citep{Guerou_2016}, NGC 3115 is an isolated massive edge-on S0 galaxy, and also the nearest S0 galaxy \citep[9.4 Mpc,][]{Arnold_2014}.
Its proximity and rich globular cluster system (GCS) make it an ideal target for studying the halo assembly history of massive early-type galaxies.

NGC 3115 is dominated by a classical bulge with a uniformly old stellar age \citep[$\gtrsim$ 13 Gyr or a formation redshift $z\gtrsim2-3$,][]{Guerou_2016}, and it hosts the nearest billion-solar-mass supermassive black hole \citep{Emsellem1999}.
It is therefore almost certain that NGC 3115 underwent a rapid early dissipative collapse, driven either by cold stream-fed gravitational instabilities \citep{Dekel2014} or gas-rich major mergers \citep[secondary-to-primary galaxy mass ratio $\geq$ 1/4,][]{Ashman1992}, leading to very high star formation and black hole accretion efficiencies and formation of the dispersion-dominated bulge. 
The high interstellar medium pressure during this early dissipative phase prompted efficient formation of massive star clusters. 
During this early epoch, major mergers for NGC 3115's primary progenitors were likely unavoidable \citep[probability $\gtrsim$ 70\%,][]{Rodriguez-Gomez2015} and essential for massive star clusters to escape early tidal disruption, migrate to the galaxy halo, and evolve into part of the present-day rich GCS, particularly the metal-rich ones \citep{Li_2004,Kruijssen2015}. 
Beyond the central bulge of NGC 3115, there exists a kinematically cold yet old ($>$ 9 Gyr) stellar disk extending to $\sim$ 4 kpc from the galactic center, which strongly suggests that no {\em disruptive} major mergers have occurred since redshift \textit{z} $\sim$ 2 \citep{Guerou_2016}. 
Similar conclusions were reached in other studies with a variety of tracers \citep{Arnold_2011,Cortesi_2013,Zanatta_2018,Poci_2019,Dolfi_2020,Buzzo_2021a}. The stellar halo mass fraction of NGC 3115 (14\%) is more than three times that of the Andromeda galaxy, despite their nearly identical stellar masses \citep{Peacock_2015}, which suggests a more extensive merger history for early-type massive galaxies like NGC 3115. The extended stellar halo of NGC 3115, including its rich GCS, may have gradually grown through minor or even mini gas-poor mergers, following the early rapid dissipative phase \citep{Dolfi_2020}. Therefore, NGC 3115 provides a clear-cut test case for the halo assembly process of massive S0 galaxies.

Mass stratification, in which more massive objects tend to concentrate toward the system’s center, is commonly observed in various gravitationally bound systems, such as stellar clusters \citep{Allison_2009,Haghi_2015,Dib_2018,Tarricq_2022} and galaxy clusters \citep{Lares_2004,Roberts_2015,Kim_2020}. However, the mass stratification of galaxy GCSs has rarely been explored. The only exceptions are the discovery of a relative deficit of massive GCs in the central regions of nearby dwarf elliptical galaxies \citep{Lotz2001}, presumably due to dynamical friction-driven inward migration and merger of massive GCs, and, more recently, evidence of GCS mass stratification in an ultra-diffuse dwarf galaxy \citep{Bar_2022}.

Previous studies on NGC 3115 have been limited to relatively small field of view, leaving the large-scale spatial distribution of its GCS largely unexplored. 
In this paper, we identify GC candidates in a large field ($70^\prime$, or $\sim 190$ kpc) around NGC 3115 and report the discovery of mass stratification in its GCS. This represents the first observational evidence of GCS mass stratification in a massive galaxy ($M_{\star}\gtrsim10^{11}\ \rm M_{\odot}$).
As we will demonstrate, the mass stratification of the GCS serves as a direct reflection of the long-sought inside-out halo assembly process of NGC 3115.
This paper is organized as follows.
The data and sample selection are described in Section \ref{sec:data}.
The analysis and results are presented in Section \ref{sec:result}.
We discuss the assembly history of NGC 3115 in Section \ref{sec:dis}.
Finally, we summarize in Section \ref{sec:sum}.

\section{Data} \label{sec:data}

DESI Legacy Imaging Surveys \citep[hereafter the Legacy Surveys,][]{Dey_2019} are motivated by the need to provide targets for the Dark Energy Spectroscopic Instrument (DESI) survey and have homogeneously mapped $\approx14,000\ {\rm deg}^2$ of the extragalactic sky.
We identify GCs on the Legacy Surveys DR9 images in \textit{g}, \textit{r}, \textit{z} bands \footnote{The images are publicly accessible on the Legacy Surveys Sky Viewer website \url{https://www.legacysurvey.org/viewer}}.
The images of the three bands are centered on NGC 3115, with the same size of $\sim153^\prime\times153^\prime$ and the same pixel scale of 0.262$^{\prime\prime}$. 

We also utilize high-precision measurements from Gaia DR3 \citep{gaia_mission_2016,gaiadr3_2023}, including proper motion, parallax, astrometric excess noise (AEN), and the blue/red photometer excess factor (BRexcess), as auxiliary data to refine the GC candidate selection.
There are 20667 Gaia sources within $70^\prime$ around NGC 3115.

\subsection{Photometry}\label{photom}

\begin{figure*}[ht]
    \centering
    \includegraphics[width=0.95\textwidth]{"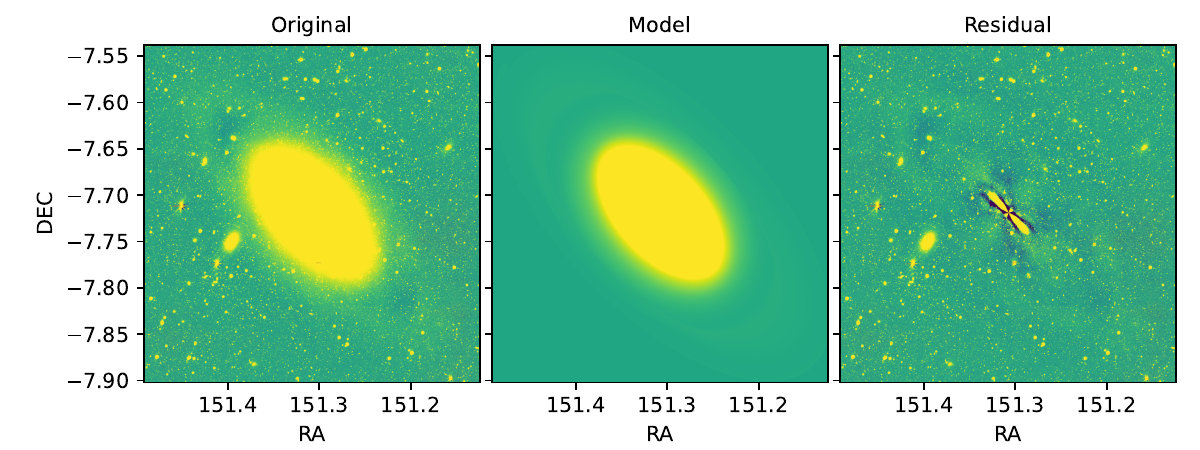"}
    \caption{The \textit{g} band images of NGC 3115.
    Left: The central region of the original \textit{g}-band image.
    Middle: The elliptical model. 
    Right: The residual image after subtracting the model.}
    \label{fig:image}
\end{figure*}
\begin{figure*}[ht]
    \centering
    \includegraphics[width=0.95\textwidth]{"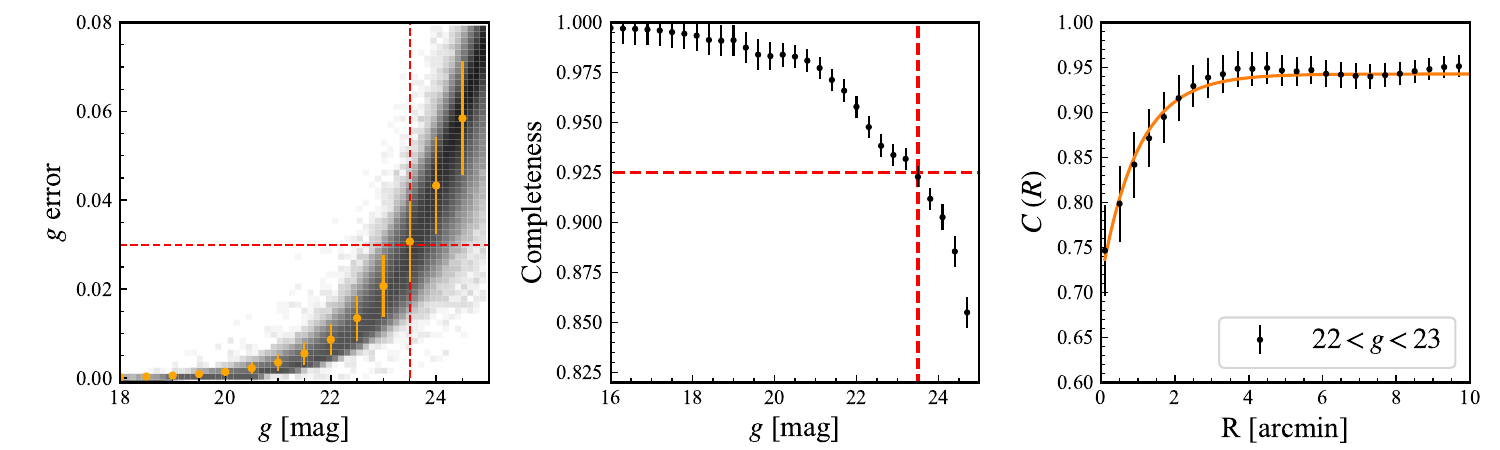"}
    \caption{Photometry results.
    Left: The \textit{g} mag error distribution. 
    Orange dots represent the mean errors in \textit{g} mag bins with errorbars representing the standard deviations.
    The vertical line marks the completeness limit of $g=23.5$, where the mean error is around 0.03 mag, marked by the horizontal line.
    Middle: The completeness as a function of \textit{g} mag.
    The error bars represent Poisson uncertainties based on the number of artificial point sources injected for the completeness test. The completeness limit of $g=23.5$ is indicated by the vertical dashed line, while the horizontal line represents the completeness of 92.5\%.
    Right: The radial completeness function for sources with $22<g<23$. The error bars represent Poisson uncertainties, and the orange curve is the exponential function fitted to the completeness in radial bins.   }
    \label{fig:photm}
\end{figure*}

The diffuse stellar light of NGC 3115 significantly impacts source detection.
Therefore, we construct an elliptical model that best fits the galaxy stellar isophotes and subtract it from the image in each band, implemented with the \textsc{photutils} module in Python \citep{photutils}.
As an example, \reffig{fig:image} displays the central region of the \textit{g}-band image, together with the ellipse model and model-subtracted residual image.
The diffuse stellar light is successfully subtracted, although the simple ellipse model struggles to adequately capture the complex central structure, 
probably including a bar \citep{Buzzo_2021a}.

We employ the dual-mode of \textsc{SExtractor} \citep{Bertin_1996} to perform detection on the \textit{g}-band residual image and perform measurement on \textit{g}, \textit{r}, \textit{z} residual images. We note that although the choice of the \textit{g} band for detection is motivated by the typically blue colors of GCs, the derived properties of the GC system are not sensitive to the specific band selected. Sources were detected above a $3\sigma$ threshold, and their magnitudes were measured using \texttt{MAG\_AUTO}, which is based on an adaptive elliptical aperture photometry method.
The errors of \textit{g} mag measurements are shown versus \textit{g} mag in the left panel of \reffig{fig:photm}, where the error is around $0.03$ mag at $g=23.5$ mag.
\texttt{MAG\_AUTO} has been shown to {\em consistently} recover $\sim$ 95\% (for its default parameter setting) of the total flux for both unresolved point sources and resolved galaxies, thereby enabling the application of consistent criteria for GC selection and fore-/background contaminant exclusion. The impact of potential spatial variations of point spread function (PSF) is automatically accounted for in this adaptive aperture photometry. We also measure some morphological parameters with \textsc{SExtractor}, including \texttt{ELLIPTICITY}, the full width at half maximum (\texttt{FWHM}), and \texttt{CLASS\_STAR}, which are useful in the candidate selection and will be detailed in Section \ref{sec:morph}.

We test the completeness of point source detection in the \textit{g}-band residual image.
First, the average PSF is extracted from the image using \textsc{photutils}, and its FWHM is $\approx 1.4$ arcsec. Considering a typical GC with effective radius of 3 pc at the distance of 9.4 Mpc, its angular radius is 0.066 arcsec, which is $<1/20$ of the PSF FWHM and well below the resolvability threshold for ground-based imaging data \citep[e.g.,][]{Chen2022}. Therefore the GCs around NGC 3115 are all point-like sources in the DESI images.
Then we inject 3000 artificial point sources each time into the image, with magnitudes randomly assigned between $16<g<25$ mag.
After that, we use {\sc SExtractor} to re-detect these artificial sources with the same parameter setup as that for the real source detection.
This process is repeated 100 times, resulting in $3\times10^5$ artificial stars.
Since the incompleteness caused by confusion is more severe in the inner region \citep{GonzalezLopezlira2017,GonzalezLopezlira2022}, especially within the central $\sim2^\prime$ affected by the residual structure, we perform an additional test for the central $6,000 \times 6,000$ pixel ($\sim 26^\prime\times 26^\prime$) region by injecting 100 artificial stars each time and repeating for 1,000 times.
In total, $4 \times 10^5$ artificial stars are used in the completeness test.
The completeness as a function of \textit{g} mag is presented in the middle panel of \reffig{fig:photm}.
The completeness limit is determined as $g=23.5$, beyond which the completeness decreases rapidly.
This limit is indicated by the dashed lines, where the completeness is around 92.5\%.
Additionally, the radial completeness function, $C(R)$, is essential for fitting the radial density profiles of GCs, which will be detailed in Section \ref{radial}.
For sources within a given magnitude range, we derive $C(R)$ by fitting the corresponding mock data points with an exponential function:
\begin{equation}
    C(R) = a \cdot \exp{(-R+b)} + c
\end{equation}
where $a,b,c$ are free parameters.
As an example, the radial completeness function for point sources with $22<g<23$ mag is shown in the right panel of \reffig{fig:photm}.
Due to incompleteness caused by source confusion, this radial completeness function drops sharply within the central $\sim4^\prime$. Beyond this radius, the function flattens and remains approximately constant at around 93\%, reflecting the overall completeness for this magnitude range. This value is also consistent with the data points shown in the middle panel. A $C(R)$ curve is derived for each magnitude bin analyzed in Section \ref{radial} and subsequently applied to correct that bin.

Although NGC 3115 lies at a relatively high Galactic latitude ($b = 36.78^\circ$), we correct for foreground extinction on a source-by-source basis using extinction coefficients and reddening maps from \citet{Schlafly2011}. The correction is implemented with the \texttt{SFDQuery} module from the \texttt{dustmaps} Python package \citep{Green2018}. 
All magnitudes and colors used hereafter have been corrected accordingly. The median extinction values across the $70^\prime$-radius field of view are $A_g = 0.15$, $A_r = 0.10$, and $A_z = 0.06$.

\subsection{GC Candidates selection} \label{selection}

\subsubsection{Fiducial sample}

We compiled a fiducial sample of spectroscopically confirmed GCs associated with NGC 3115 as a reference for determining the optimal criteria for selecting GC candidates.
The first sample is from the SLUGGS Survey \citep{Brodie_2014}, which used spectroscopy from Keck/DEIMOS and provided 150 confirmed GCs. 
The second sample is from \citet{Jennings_2014}, who identified 781 photometric candidates based on HST/ACS and Subaru/Suprime-Cam, but only 176 out of the 781 candidates have radial velocity measurements between $350\sim1200$ km/s.
Cross-matching these two samples with a 1$^{\prime\prime}$ matching radius results in 191 distinct confirmed GCs, 160 of which have counterparts in our DESI-Gaia catalog.

Besides, Gaia provides a catalog of galaxy candidates \citep{gaia_extragalactic_2023}. Its classification favors completeness over purity, thus some galaxy candidates are actually misclassified stars. 
These stars will be removed by Gaia astrometric parameters (see Section \ref{astrom}), and remaining sources will serve as a reference galaxy sample.

\subsubsection{Astrometric parameters}\label{astrom}

We first utilize the high-precision measurements of proper motion and parallax from Gaia.
At the 9.4 Mpc distance of NGC 3115, GCs are too distant to exhibit observable proper motion or parallax \citep{barmby_gaia_2022,gaia_collaboration_crf3_2022},
thus any source with significant parallax or proper motion is classified as a foreground star.
We reduce these two parameters with their respective errors \citep{barmby_gaia_2022,gaia_collaboration_crf3_2022}:
\begin{equation}
    \varpi^*\equiv\left|{\frac{\varpi+0.017}{\sigma_\varpi}}\right|
\end{equation}
\begin{equation}
    \mu^*\equiv 
        \sqrt{
        \left[\mu_\alpha\ \mu_\delta \right]{\rm Cov}(\mathbf{\mu})^{-1} 
        \left[
            \begin{array}{c}   
                \mu_\alpha\\  
                \mu_\delta\\  
            \end{array}
        \right]      }
\end{equation}
where $\varpi$ and $\sigma_\varpi$ represent the parallax and its error, 0.017 milliarcseconds is the median parallax zero point correction \citep{Lindegren_2021}, $\mu_\alpha$ and $\mu_\delta$ denote two proper motion components, and ${\rm Cov}(\mathbf{\mu})$ is their covariance matrix \citep{gaia_collaboration_crf3_2022}.
A source is classified as a foreground star at a significance of $3\sigma$ if $\varpi^*>3$ or $\mu^*>3$.

\subsubsection{Excess Factors}

\begin{figure*}[ht]
    \centering
    \includegraphics[width=0.95\textwidth]{"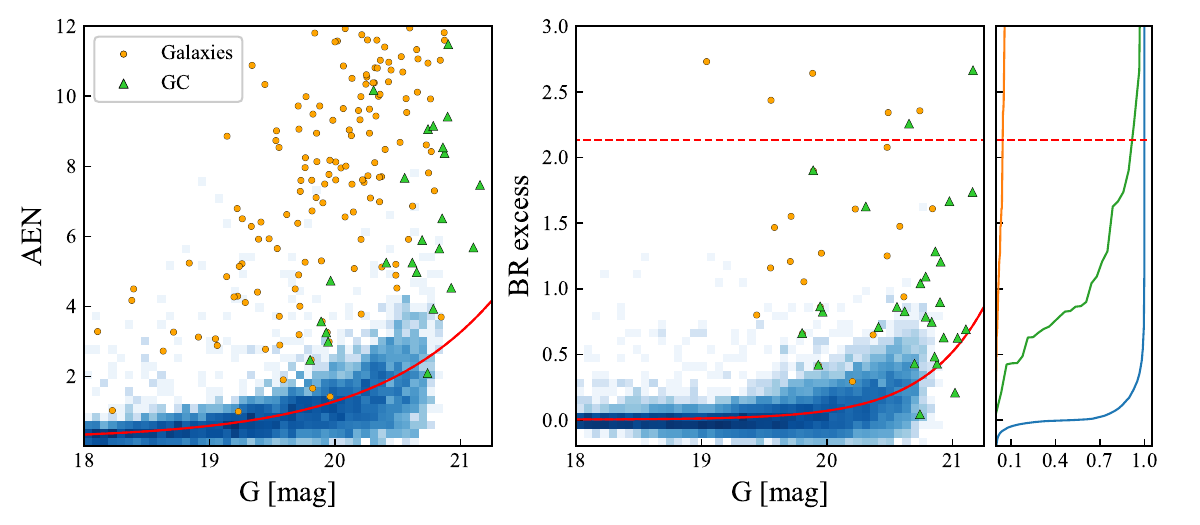"}
    \caption{Distributions of excess factors.
    Left: AEN against \textit{g} mag for stars (blue shading), galaxies (orange dots), and GCs (green triangles).
    The red solid curve represents the threshold for GC selecting candidates.
    Middle: similar to the left but for BRexcess. 
    The red dashed horizontal line represents the criterion of BRexcess $=2.5$ for excluding galaxies.
    Right: the cumulative BRexcess distributions of stars (blue), galaxies (orange), and GCs (green).}
    \label{ExcessFactor}
\end{figure*}

Two excess factors provided by Gaia, Astrometric Excess Noise (AEN) and Blue Photometer/Red Photometer Excess Factor (BRexcess), have proven effective in identifying extended sources, including GCs \citep{voggel_2020,hughes_2021,Pan_2022,Obasi_2023,Obasi_2024}.
AEN represents the quality of the astrometric five-parameter model fitting in Gaia, which assumes standard point sources.
Thus extended objects are typically poorly fitted and have AEN $>0$ \citep{lindegren_2018,gaia_collaboration_2018}. 
BRexcess is defined as the ratio of the summed flux in the Blue Photometer 
(BP, $3300\sim 6800$ \AA) and the Red Photometer (RP, $6400\sim 10500$ \AA) to the total flux in the broad \textit{g} band ($3300\sim 10500$ \AA) \citep{evans_2018}. 
The \textit{g} band flux is obtained by spectral profile fitting assuming point sources, while BP/RP flux is measured directly from a square aperture. 
Thus for an extended source, its BP/RP flux is accurate while its \textit{g} band flux is underestimated, resulting in an elevated BRexcess value.
Moreover, the original BRexcess values from Gaia DR3 exhibit a significant correlation with the $\rm BP-RP$ color, potentially introducing selection bias. 
We corrected this bias using a color-dependent term provided by \citet{riello_2021}.

The stars selected by parallax and proper motion are used here to determine optimal thresholds for distinguishing GCs from stars, as shown in \reffig{ExcessFactor}.
Since both AEN and BRexcess exhibit significant dependence on Gaia $G$ mag, we construct thresholds varying with $G$ mag by binning the stars with widths of 0.1 mag and fitting an exponential function to the mean values in these bins:
\begin{equation}
    {\rm AEN} = 1.13\times10^{-9}\times e^{1.03G}+0.21
\end{equation}
\begin{equation}
    {\rm BRexcess}= 8.62\times10^{-19}\times e^{1.95G}+0.005
\end{equation}
These two functions are indicated as the red curves in \reffig{ExcessFactor}, and sources below them are classified as foreground stars.

Additionally, BRexcess is effective for excluding galaxies.
The right panel of \reffig{ExcessFactor} displays the cumulative BRexcess distributions for galaxies, GCs, and stars.
We set the 95th percentile of GC distribution (BRexcess $=2.5$) as the threshold, and it can exclude over 90\% of galaxies while retaining 95\% of the fiducial GCs.

\subsubsection{Morphological Parameters}
\label{sec:morph}

\begin{figure*}[ht]
    \centering
    \includegraphics[width=0.95\textwidth]{"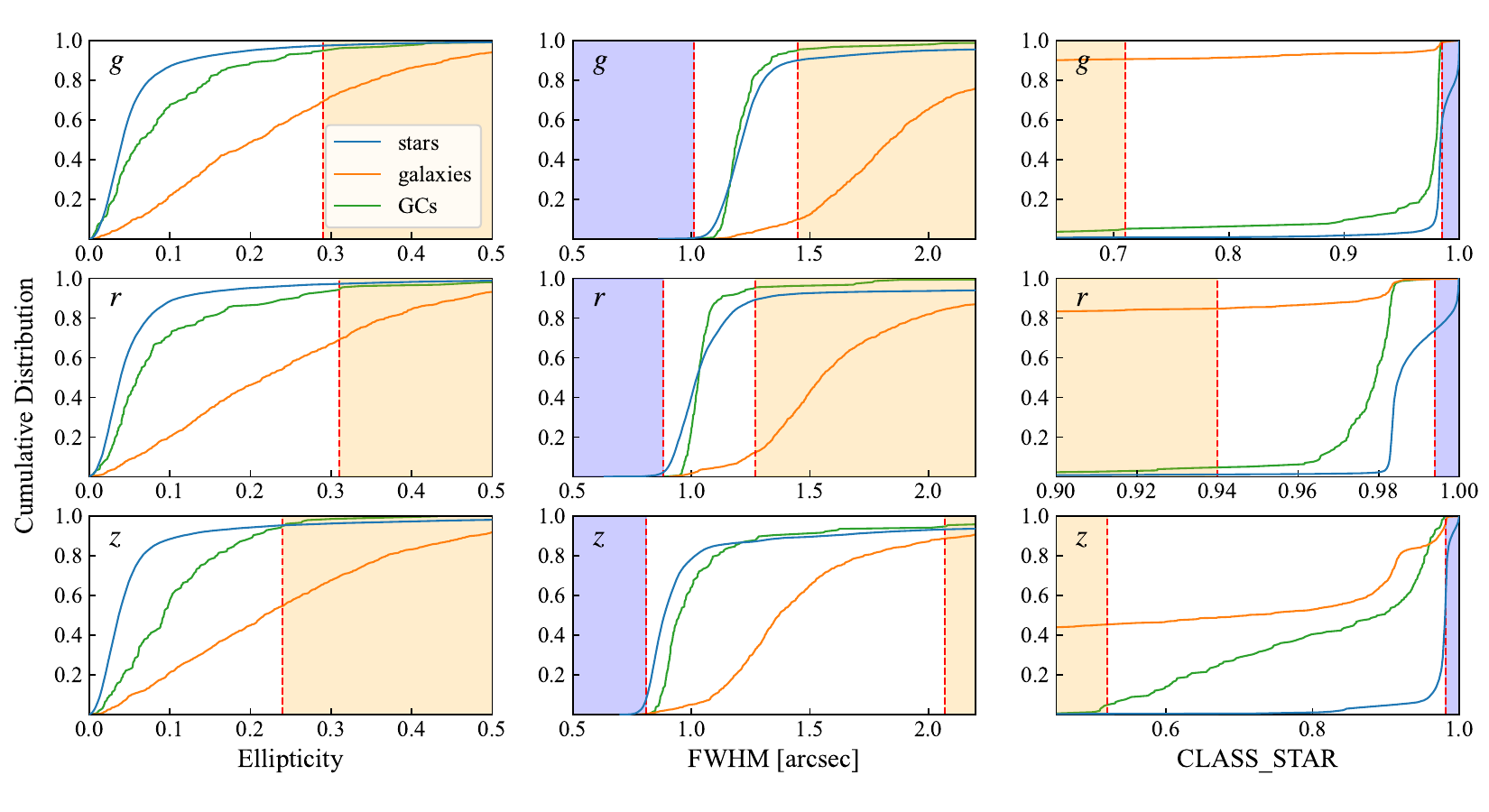"}
    \caption{The morphological parameters.
    The cumulative distributions of \texttt{ELLIPTICITY}, \texttt{FWHM}, and \texttt{CLASS\_STAR} are shown in the left, middle, and right columns, respectively. The results for the \textit{g}, \textit{r}, and \textit{z} band are displayed in the top, middle, and bottom rows, respectively.
    The red dashed vertical lines in each panel represent the criteria for selecting GC candidates, and the blue and orange shaded regions represent the excluded ranges for stars and galaxies, respectively.}
    \label{desi_morph}
\end{figure*}

Besides Gaia measurements, there are three widely adopted morphological parameters measured from the DESI images using \textsc{SExtractor}, which can be used to select GC candidates: \texttt{ELLIPTICITY}, \texttt{FWHM} and \texttt{CLASS\_STAR}. 
The cumulative distributions of these parameters in different bands are shown in \reffig{desi_morph}, where colors indicate three types of objects: stars (blue), GCs (green), and galaxies (oranges).
The star and galaxy samples are defined based on Gaia parameters as described earlier: stars are identified as sources with significant astrometric measurements or with excess factors below the critical threshold, while galaxies are either Gaia galaxy candidates without significant proper motion or parallax, or sources with BRexcess values greater than 2.5.
Note that the \texttt{ELLIPTICITY} parameter is measured only on the detection image in the dual-mode of SExtractor, which, in our case, corresponds to the g-band image. 
To obtain ellipticity measurements in the r- and z-bands, we additionally run \textsc{SExtractor} using those bands as detection images.
However, all other photometric measurements are performed in dual-mode described in Section~\ref{photom}.

\texttt{ELLIPTICITY} and \texttt{FWHM} quantify the elongation and angular size of a source, respectively. Since GCs and stars are expected to be nearly circular and point-like, they are largely indistinguishable in these parameters. However, sources with elongated shape or extended sizes are almost certainly galaxies.
Therefore, we adopt the 0th and 95th percentiles of GCs distributions as thresholds for \texttt{ELLIPTICITY} and \texttt{FWHM}, which are
$0<$ \texttt{ELLIPTICITY}\_\textit{g} $<0.29$, 
$0<$ \texttt{ELLIPTICITY}\_\textit{r} $<0.31$, 
$0<$ \texttt{ELLIPTICITY}\_\textit{z} $<0.24$, 
$1.01<$ \texttt{FWHM}\_\textit{g} $<1.45$, 
$0.88<$ \texttt{FWHM}\_\textit{r} $<1.27$, and 
$0.81<$ \texttt{FWHM}\_\textit{z} $<2.07$.
These thresholds, along with the corresponding ranges for stars and galaxies, are illustrated in \reffig{desi_morph} using red dashed lines, blue shaded regions, and orange shaded regions, respectively.
This morphological filtering removes only a small fraction of stars but eliminates approximately 90\% of galaxies.

The \texttt{CLASS\_STAR} parameter is based on a multilayer feed-forward neural network and serves as a classifier for distinguishing between point-like ($\texttt{CLASS\_STAR}=1$) and extended sources ($\texttt{CLASS\_STAR}=0$).
Similarly, we adopt the 5th and 100th percentiles of GCs distribution as selection thresholds for \texttt{CLASS\_STAR}: 
$0.710<$ \texttt{CLASS\_STAR}\_\textit{g} $<0.985$, 
$0.940<$ \texttt{CLASS\_STAR}\_\textit{r} $<0.994$, and 
$0.520<$ \texttt{CLASS\_STAR}\_\textit{z} $<0.981$.
As shown in the right panels of \reffig{desi_morph}, more than $20\%$ of stars and $90\%$ of galaxies are removed by these criteria.

\subsubsection{Color}

\begin{figure*}[ht]
    \centering
    \includegraphics[width=0.95\textwidth]{"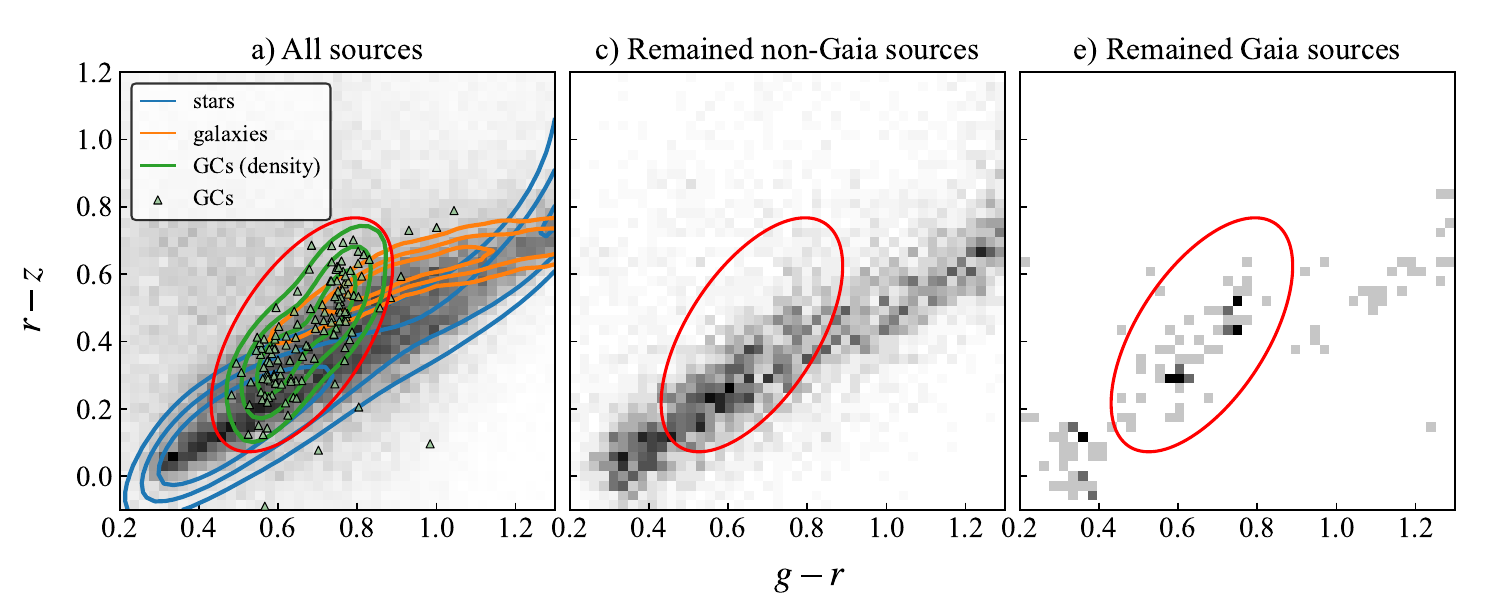"}
    \caption{The $g-r$ vs. $r-z$ diagram.
    The left panel presents all sources as the gray shading. The distributions of stars and galaxies are represented respectively as blue and orange contours.
    Individual fiducial GCs are indicated as green triangles.
    The middle and right show the remaining non-Gaia and Gaia sources after previous selection, respectively.
    The red ellipse represents the GC region.}
    \label{color}
\end{figure*}
The last selection criterion is based on the $g-r$ vs. $r-z$ distribution, as shown in \reffig{color}.
In the left panel, the 30\%, 60\%, and 90\% number density contours of stars, GCs, and galaxies, constructed using kernel density estimation (KDE), are overlaid on the distribution of all sources.
Here, the star and galaxy samples are the same as those in Section \ref{sec:morph}, while GCs are the fiducial sample. 
Stars, GCs, and galaxies occupy distinct regions in color space, although with some degree of overlap. 
We define an elliptical region to characterize the locus of GCs, centered at $g-r=0.66$ and $r-z=0.42$, with a semimajor axis of 0.38 mag, a semiminor axis of 0.17 mag, and a position angle of $63^\circ$ measured from the $g-r$ axis.
The middle and right panels show the remaining sources after applying all but the color selection criteria. 
Noticeable overdensities are evident in the elliptical region occupied by GCs, especially for sources with Gaia measurements.

\subsubsection{GC candidate sample}

The final sample of GC candidates includes sources that meet all the above criteria, lie within the elliptical GC region in the color-color space, and are brighter than the completeness limit of $g=23.5$. 
This yields a total of 2193 GC candidates within a $70^\prime$ ($\sim 190$ kpc) radius around NGC 3115, with \textit{g}-band magnitude ranging from 19.3 to 23.5 mag.
This sample is presented in Table \ref{tab:cat} and used in the subsequent analysis presented in Section \ref{sec:result}.
Given the stellar effective radius of $R_{\rm e,\star}=57^{\prime\prime}$ for NGC 3115 \citep{Capaccioli_1987}, our $70^\prime$-radius detection extends to $\sim74 R_{\rm e,\star}$, corresponding to $\sim$ 1.1 times the virial radius of NGC 3115, based on the average scaling relation from \cite{Kravtsov2013}.

It is important to note that contaminants dominate the sample in the outer regions.
As shown in Section \ref{radial}, the number density of GC candidates begins to flatten to the background level beyond $30^\prime$ (∼82 kpc), within which there are 731 GC candidates prior to background subtraction. 
We also note that 75\% of fiducial GCs are successfully recovered by our selection criteria.
Whenever relevant in the following analysis, we will compensate for the fraction (25\%) of GCs missed by our selection method.

\begin{table*}
    \centering
    \caption{Catalog of 2193 GC candidates.}
    \label{tab:cat}
\begin{threeparttable}
\begin{tabularx}{1\textwidth}{cccccccccc}
\toprule
ID & ra & dec & R\tnote{a} & gmag\tnote{b} & gerr\tnote{b} & ellip\_g\tnote{c} & fwhm\_g & CLASS\_STAR\_g & gaia\_id \\
   & [deg] & [deg] & [arcmin] & & & & [arcsec] & & \\
\midrule
1 & 151.321448 & -7.722002 & 0.83 & 20.5638 & 0.0018 & 0.0876 & 1.2542 & 0.9804 & 3772282699634372224 \\
2 & 151.305557 & -7.704874 & 0.84 & 20.9150 & 0.0023 & 0.1811 & 1.2646 & 0.9822 & 3772282974511877120 \\
3 & 151.304938 & -7.704780 & 0.85 & 20.1464 & 0.0011 & 0.0896 & 1.2120 & 0.9830 & 3772282974511780736 \\
4 & 151.304562 & -7.732748 & 0.87 & 19.8273 & 0.0011 & 0.1710 & 1.2143 & 0.9570 & 3772282562195640064 \\
5 & 151.297076 & -7.705924 & 1.00 & 20.6301 & 0.0021 & 0.0996 & 1.2382 & 0.9801 & 3772282871432566656 \\
\bottomrule
\end{tabularx}
\begin{tablenotes}
    \footnotesize
    \item[*] Table 1 is published in its entirety in the machine-readable format.
    Only a portion of the table and the parameters in \textit{g} band are shown here for guidance regarding its form and content.
    \item[a] The galactocentric radius.
    \item[b] \textit{g} band magnitudes and their errors. 
    \item[c] \textit{g} band ellipticity.
\end{tablenotes}
\end{threeparttable}
\end{table*}

\subsection{TNG50 simulations}\label{tng50}

As we will show in the discussion section, the distribution of GCs in NGC 3115 encodes the assembly history of the stellar halo. Here we briefly describe the numerical simulations to be used for decoding the halo assembly based on the luminosity- or mass-dependent spatial distribution of GC candidates in NGC 3115.

We utilize the TNG50-1 simulation (hereafter TNG50) to decode the assembly histories of NGC 3115 analogues.
TNG50 is the highest-resolution flagship run of the IllustrisTNG cosmological magnetohydrodynamical simulations of galaxy formation within the $\Lambda$CDM framework \citep{Nelson2019,Pillepich2019}. It simulates the evolution of dark matter, gas, stars, and black holes within a periodic cubic volume of 51.7 comoving Mpc per side, achieving a baryonic mass resolution of 8.5$\times$10$^{4}$ $M_{\odot}$ and an average spatial resolution (cell size) of 70–140 pc. The unprecedented resolution of TNG50 makes it particularly well-suited for studying the assembly history of massive galaxies in a cosmological context. 

\subsubsection{Selection of NGC 3115-like subhalos}

To select subhalos that have structures and formation history resembling those of NGC 3115, we begin by selecting primary subhalos (central galaxies) at redshift $z=0$ with stellar masses in the range of  $10^{10.9}$$-10^{11.3}\rm\ M_{\odot}$ and bulge-to-total stellar mass ratio of $f_{\rm bulge}>0.6$, where $f_{\rm bulge}$ is defined as twice the fractional mass of stars with circularity parameter $\epsilon<0$ \citep{Genel2015}. 
As a reference, the bulge-to-total stellar mass ratio of NGC 3115 was estimated to be $f_{\rm bulge}=0.8$ \citep{Bell2017}.
Additionally, given the existence of an old, thin, and fast-rotating stellar disc, dominated by stars that formed some 9 Gyr ago, with a small contribution of younger stars \citep{Norris2006,Guerou_2016,Poci_2019}, \citet{Dolfi_2020} proposed that NGC 3115 has undergone an early gas-rich merger of mass-ratio around 1:4-1:10 9 Gyr ago and evolved passively after that. Therefore, we exclude subhalos that have undergone major mergers (secondary-to-primary stellar mass ratio $\mu$ $>$ 1/4) since $z < 1$, and require that the fractional mass contribution from {\em in situ} star formation since $z < 1$ is less than 5\%.
We use the {\sc SubLink} algorithm \citep{Rodriguez-Gomez2015} to trace merger trees and identify main-branch progenitors and sub-branch progenitors. The stellar assembly catalog of \citet{Rodriguez-Gomez2016} is used to extract the stellar mass formed {\em in situ} or {\em ex situ}. Based on the above criteria, we identify five primary subhalos that match the main properties of NGC 3115.

\subsubsection{Probabilistic assignment of GCs to individual subhalos}\label{sec:gctag_tng50}

There exists a general correlation between the number of GCs and the stellar mass of their host galaxies \citep[e.g.,][]{Harris1981,Peng2008,Zaritsky_2015,Lim2018,Liu2019}. 
However, the scatter in this correlation increases toward both the low- and high-mass ends, leading to a characteristic V-shaped average relation between the specific frequency of GCs ($S_{\rm N}$: a quantification of the richness of GCs relative to the host galaxy luminosity) and the stellar mass of the host galaxies \citep[e.g.,][]{Peng2008,Lim2018,Liu2019,Choksi2019}. We adopt the observed $S_{\rm N}$ distribution as a function of the host galaxy’s $V$-band magnitude ($M_{V}$) for normal galaxies \citep{Lim2018}. Specifically, we assume a uniform $S_{\rm N}$ distribution with a lower limit of zero and an upper limit defined by the 1-$\sigma$ upper envelope of the distribution as a function of $M_{V}$ for normal galaxies in the Coma cluster \citep{denBrok2014}. This distribution broadly encompasses galaxies from various environments, including those in the Virgo and Fornax clusters, as well as field galaxies. For galaxies outside the $M_{V}$ ranges (from $-13$ to $-17$ mag) covered by observations, we assume the same $S_{\rm N}$ distribution as at the lower (for $M_{V}>-13$ mag) or higher (for $M_{V}<-17$ mag) luminosity ends.

The observed $M_{V}$$-$$S_{\rm N}$ relation for nearby galaxies cannot be directly applied to simulated subhalos at arbitrary cosmic epochs, as the stellar mass-to-luminosity ratio evolves over time. To address this, we apply the above $M_{V}$$-$$S_{\rm N}$ relation to TNG50 subhalos at $z = 0$ by randomly sampling the $S_{\rm N}$ distribution within the $M_{V}$-dependent lower and upper bounds. The resulting distribution is then sorted by galaxy stellar mass. This yields a stellar mass-dependent distribution of the number of GCs per unit stellar mass, $p_{M}$, which can then be used to probabilistically assign GCs to individual subhalos.

\subsubsection{GC-weighted average stellar mass of merged galaxies}\label{sec:avmass_tng50}

We trace all star particles in the five TNG50 subhalos back along the merger tree to identify their parent galaxies at earlier epochs. Star particles whose parent galaxies lie on the main branch are classified as {\em in situ}, with the parent’s stellar mass taken as the subhalo’s stellar mass at $z = 0$. Those originating from sub-branch parent galaxies are classified as {\em ex situ}, and their parent’s stellar mass at infall is recorded. 

Using the parent galaxy stellar mass $M_{i, \star}$ associated with each star particle $i$, along with the mass-dependent probabilistic GC contribution $p_{i, M}$, we compute the GC-weighted average stellar mass of merged galaxies at different galactocentric distances as:
\begin{equation}
   {\langle M_{\star, {\rm parent}}\rangle} = \frac
   {\sum_{i} p_{i, M} \cdot M_{i,\star}}
   {\sum_i p_{i, M}}
\end{equation}
This can be directly compared to the average parent galaxy mass estimated from observations, as will be detailed in Section \ref{decode}. 

Lastly, we note that a post-processing catalog of GCs tagged to dark matter particles in the most massive TNG50 groups and clusters was presented by \cite{Doppel2023}. Two of our five NGC 3115-like subhalos are included in their sample. However, it turns out each of the two subhalos hosts only about 60 GCs—substantially fewer than the expected total number for NGC 3115. Based on our star-particle tagging method described above, the five NGC 3115 analogs have average GC numbers ranging from 228 to 405. Therefore, we opt to use our star particle-tagged probabilistic method to explore the GC-weighted average host galaxy mass of the subhalos. This is sufficient for our exploratory investigation. A more sophisticated analysis, explicitly taking into account the GC luminosity distribution as a function of galactocentric radius, will be presented in future work.


\section{Analysis and Results}\label{sec:result}

\subsection{Radial density profiles}\label{radial}

\begin{figure}
    \centering
    \includegraphics[width=0.95\linewidth]{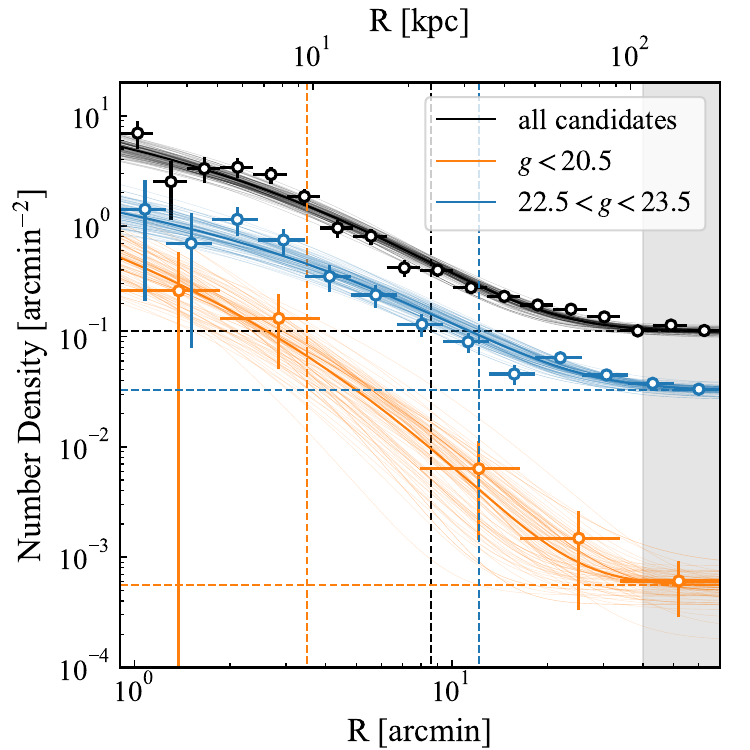}
    \caption{The Sersic radial density profiles of the entire (black), the brightest (orange), and the faintest (blue) GC candidate samples.
    The data points are only for visual inspection, with error bars representing Poisson errors. 
    The vertical and horizontal lines indicate the effective radii and the background levels, respectively. 
    The light-colored curves show 200 random MCMC samplings that illustrate the fitting uncertainties.
    The gray shading indicates the region between $40^\prime$ and $70^\prime$ used for background estimation.} 
    \label{Profile_eg}
\end{figure}

\begin{figure*}
    \centering
    \includegraphics[width=0.95\linewidth]{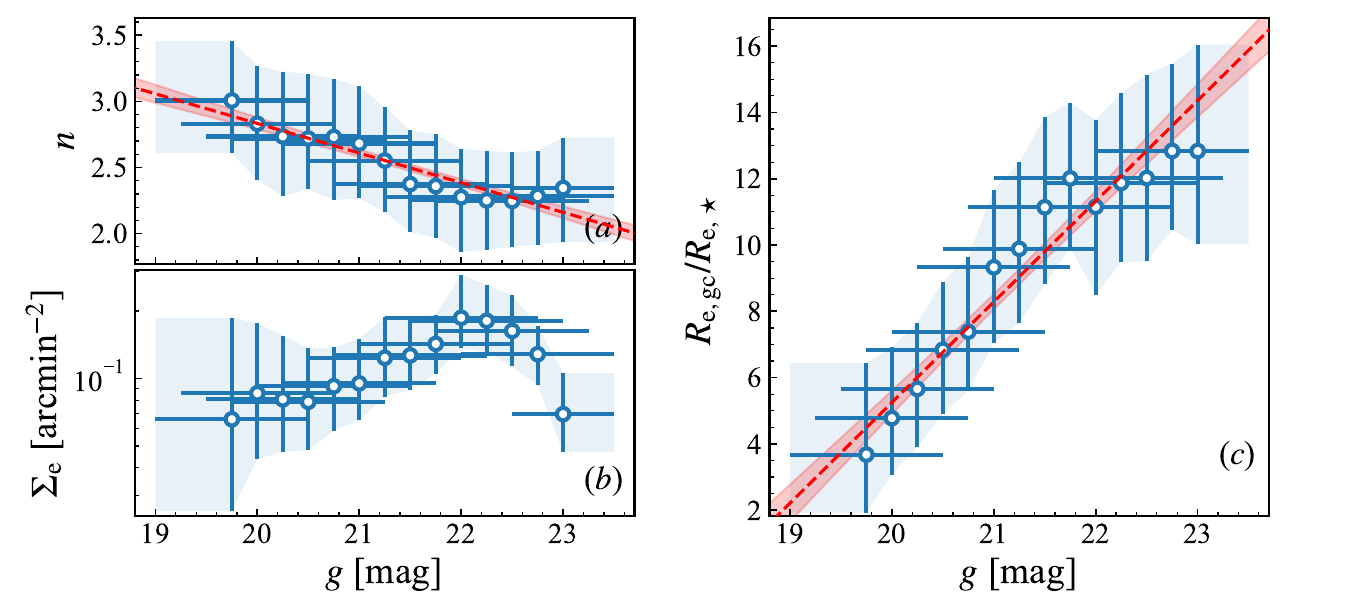}
    \caption{Sersic profile parameters as functions of \textit{g} mag.
    Panel (a) shows the Sersic index $n$, panel (b) shows the surface density at the effective radius $\Sigma_{\rm e}$, and panel (c) shows the effective radius of the GC system normalized by the stellar effective radius, $R_{\rm e,gc}/R_{\rm e,\star}$.
    Vertical error bars represent the 16th and 84th percentiles of the MCMC sampling, while horizontal error bars indicate the magnitude range of each bin.
    Red dashed lines in panel ($a$) and ($c$) indicate the best-fit linear trends in $n$ and $R_{\rm e,gc}/R_{\rm e,\star}$, respectively.  }
    \label{profile_params}
\end{figure*}

We quantify the spatial distribution of GC candidates across different \textit{g}-band magnitudes by fitting their radial density profiles. To ensure an adequate number of candidates in each sample, the \textit{g} mag bins are defined with a fixed width of 1.5 mag and center positions spaced by 0.25 mag.
The only exception is the faintest bin, which is centered at $g = 23$ mag with a width of 1.0 mag.

We employ the Sersic function to fit the radial profiles:
\begin{equation}
    \Sigma_{\rm sersic}(r) = \Sigma_{\rm e} \exp\left\{-b_n\left[ \left(\frac{r}{R_{\rm e}}\right)^{1/n}-1 \right]\right\} + \Sigma_{\rm b}
\end{equation}
where $b_n$ is a function of the Sersic index $n$, with specific formula given by \citet{Ciotti_1999}.
$\Sigma_e$ is the density at the effective radius, $R_{\rm e}$.
The constant term $\Sigma_b$ represents the average background (or contaminant) number density, contributed by foreground stars and distant galaxies. The average background is evident (and thus well constrained) from the flattening of the GC candidate radial profile at large galactocentric radii.
The Sersic models are fitted to individual candidates using Markov Chain Monte Carlo (MCMC) method, implemented with \texttt{emcee} in Python \citep{emcee}.
The likelihood function is defined as:
\begin{equation}
    \mathcal{L} \propto \prod_i l_i(r_i|\Sigma_{\rm sersic})\cdot {\rm C}(r)
\end{equation}
where $l_i$ represents the probability of finding the $i$ th data point at radius $r_i$ given the density profile $\Sigma_{\rm sersic}$, and ${\rm C}(r)$ represents the radial completeness function described in Section \ref{photom}.

We apply flat priors to the parameters: $0 < n < 5$, $0.1^\prime < R_{\rm e} < 20^\prime$, and $0.01\ \mathrm{arcmin}^{-2} < \Sigma_{\rm e} < 1\ \mathrm{arcmin}^{-2}$.
A Gaussian prior is set to the background density $\Sigma_{\rm b}$.
Its mean and standard deviation are estimated with GC candidates located within an annular region between $40^\prime$ and $70^\prime$, through bootstrapping along the position angle for 100 iterations and each time randomly taking an arc of $60^\prime$. 

As examples, the radial profiles of the entire (black), the brightest (orange), and the faintest (blue) GC candidate sample are presented in \reffig{Profile_eg}, with light-colored curves indicating 200 random MCMC samplings.
Note that the data points are shown solely for visualization purposes and the fitting is performed on individual GC candidates. 

The fitted parameters of Sersic profiles are plotted against \textit{g} mag in \reffig{profile_params}.
As \textit{g} mag increases, the Sersic index ($n$, panel \textit{a}) shows a weakly decreasing trend, indicating flatter profiles for fainter ones in general.
The number density at the effective radius ($\Sigma_{\rm e}$, panel \textit{b}) reaches a maximum at around $g=22$ and decreases toward both the brighter and fainter end.
Most notably, the systematical variation in radial distributions is directly reflected by the effective radius ($R_{\rm e,gc}$, panel \textit{c}), which increases systematically with \textit{g} mag.
It suggests that brighter GCs are more centrally concentrated, whereas fainter ones exhibit more extended distributions.
The radial variation in both $n$ and $R_{\rm e,gc}$ approximately follow linear relations:
\begin{equation} \label{equ:n_g}
    n = ( -0.22\pm0.03)\times g+(7.31\pm0.54) 
\end{equation}
\begin{equation} \label{equ:Re_g}
    R_{{\rm e,gc}}/R_{{\rm e,\star}} = ( 3.04\pm0.24)\times g+(-55.49\pm5.04) 
\end{equation}
where $R_{\rm e,\star}=57^{\prime\prime}$ is the stellar effective radius of NGC 3115 \citep{Capaccioli_1987}.
While the \textit{g}-band stellar mass-to-light ratio increases with optical colors, its overall variation is less than 0.2 dex for classical GCs \citep{Jordan_2007}. 
Taken together, these trends provide strong evidence of mass stratification in the GCS of NGC 3115, with lower-mass GCs extending to larger galactocentric radii than their higher-mass counterparts.

\subsection{Globular cluster luminosity functions}\label{gclf}

The globular cluster luminosity function (GCLF) is one of the fundamental properties of GCSs.
It is widely known that GCLF has a similar shape across different galaxies \citep[e.g.,][]{Harris1996,Bica2006,Peacock2010}, which is typically modeled with a Gaussian function \citep{Hanes_1977,Secker_1992,Villegas_2010,Brodie_2012,Harris_2014,Lomeli2022}, characterized by a nearly constant turnover magnitude and a standard deviation (or dispersion) that varies with the luminosity of host galaxy \citep{Jordan_2007}.
Here, we measure both the overall GCLF of the entire GCS and GCLFs at various galactocentric radii.

\subsubsection{Overall GCLF}\label{gclf_total}

\begin{figure}
    \centering
    \includegraphics[width=0.95\linewidth]{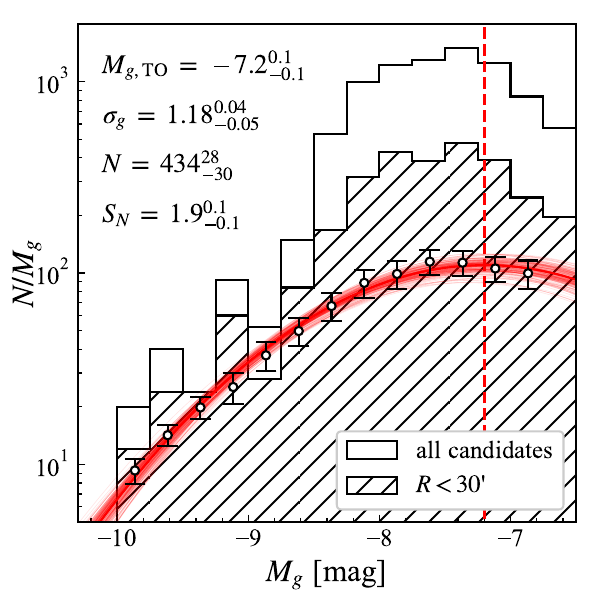}
    \caption{The \textit{g}-band GCLF of the entire GCS.
    The open and hatched histograms show the raw distribution of GC candidates (without background subtraction) within 70$^{\prime}$ and the central 30$^{\prime}$, respectively.
    The black dots with error bars, representing the background-subtracted GC number in the same \textit{g} mag bins as in Section \ref{radial}, are used for fitting the Gaussian GCLF. 
    The light red curves represent Gaussian fittings for 200 random MCMC sampling of the measurements, illustrating the fitting uncertainty.
    The fitted GCLF parameters, total number of genuine GCs, and the corresponding specific frequency of GCs in NGC 3115 are given in the upper left corner.
    The best-fit turnover magnitude is also indicated as the dashed vertical line. See Section \ref{gclf_total} for more details. }
    \label{GCLF_total}
\end{figure}

To construct the \textit{g} band GCLF of the entire GCS of NGC 3115, we first estimate the genuine GC numbers in various \textit{g} mag bins by integrating the magnitude-dependent Sersic profiles obtained in Section \ref{radial}, over the radial range from $0^\prime$ to $70^\prime$, after subtracting the background. 
The estimated numbers of genuine GCs are indicated as data points in \reffig{GCLF_total}. We employ MCMC to fit a standard Gaussian function to the estimated numbers of genuine GCs across different $g$-mag bins:
\begin{equation}
    N(M_g|M_{g,\rm TO},\sigma_g) = N_0\cdot e^{-\frac{(M_g-M_{g,\rm TO})^2}{2\sigma_g^2}}
\end{equation}
where $M_g$ represents the \textit{g}-band absolute magnitude. 
The parameters for fitting are turnover magnitude $M_{g,\rm TO}$ and dispersion $\sigma_g$, to which we apply Gaussian priors according to the typical values obtained by \citet{Jordan_2007}: $\mathcal{N}(-7.2, 0.2^2)$ for $M_{\rm g,TO}$ and $\mathcal{N}(1.14, 0.1^2)$ for $\sigma_g$.

The fitting results are shown in \reffig{GCLF_total}, where the open and hatched histograms represent the raw counts (i.e., without background subtraction) of all GC candidates and candidates within 30$^{\prime}$, respectively. 
The light red curves represent Gaussian fitting from 200 random samplings of the measurements.
The fitted parameters are given in the upper left corner: $M_{g,\rm TO}=-7.2_{-0.1}^{+0.1}$ and $\sigma_g=1.18_{-0.05}^{+0.04}$.
The expected total number of genuine GCs of NGC 3115 is estimated by integrating the Gaussian GCLF.
Considering that $25\%$ of genuine GCs are expected to be misclassified as contaminants in the selection process (Section \ref{selection}), 
we multiply the total number of genuine GCs by 1/0.75 to account for this incompleteness.
This yields a final count of $N = 434_{-30}^{+28}$, corresponding to a specific frequency of $S_N=1.9^{0.1}_{-0.1}$.
This result is in good agreement with previous studies: \citet{Hanes_1986} identified a total of $N=520\pm120$ GCs through star counts upon CFHT and AAT prime focus plates, \citet{Harris2011} reported $N=550\pm150$ and a specific frequency of $S_N=2$, and \citet{Faifer_2011} obtained $N=571\pm190$ GCs with $S_N=2\pm0.7$ using Gemini/GMOS images.
In addition, since candidates beyond $30^\prime$ are dominated by contaminants, we estimate a sample purity of $\sim60\pm4\%$ by comparing the number of genuine GCs ($434\pm29$) to the total number of candidates within $30^\prime$ (731).

\subsubsection{GCLF at various radii}\label{gclf_radial}

\begin{figure*}
    \centering
    \includegraphics[width=0.95\linewidth]{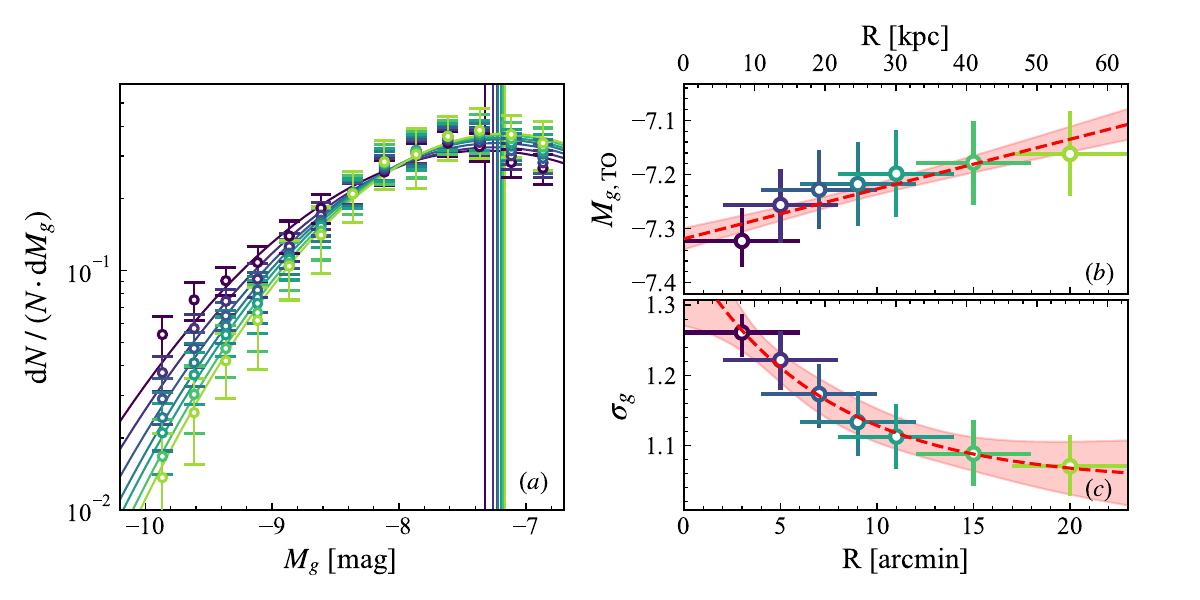}
    \caption{The radial variation in \textit{g}-band GCLF.
     Panel ({\em a}) displays GCLFs in various radial bins. The color scheme is such that lighter colors correspond to larger radii. The vertical lines indicate the turnover magnitude for different radial bins. 
     Panel ({\em b}) and panel ({\em c}) show the radial variations of turnover magnitude ($M_{g,\rm TO}$) and dispersion ($\sigma_g$), respectively.
     The horizontal error bars represent the radial ranges for corresponding bins, and vertical error bars represent the 16th and 84th percentiles of the MCMC samplings.
     The red dashed line and curve represent the best-fit relations for $M_{g,\rm TO}$ and $\sigma_g$, respectively.}
    \label{GCLF_R}
\end{figure*}

The mass stratification revealed by radial profiles in Section \ref{radial} is also evident in the variation of GCLF across different galactocentric radii.

We first divide the candidates into seven radial bins, centered at $3^\prime$, $5^\prime$, $7^\prime$, $9^\prime$, $11^\prime$, $15^\prime$ and $20^\prime$, each with a width of $3^\prime$.
These bins partially overlap with the next ones to ensure sufficient large subsamples for fitting.
For each radial bin, we construct the corresponding GCLF through the same method demonstrated in Section \ref{gclf_total}.
The data points representing true GC numbers and the fitted curves representing GCLFs are all shown in panel ({\em a}) of \reffig{GCLF_R}.
Their colors represent different radial bins, with lighter colors corresponding to larger radii.

The fitted \textit{g}-band parameters, $M_{g,\rm TO}$ and $\sigma_g$, are plotted against galactocentric radius in panel ({\em b}) and ({\em c}), respectively.
$M_{g,\rm TO}$ generally increases with radius, while $\sigma_g$ exhibits a piece-wise declining trend: it decreases sharply toward $R \sim 10^\prime$, beyond which the decline slows down.
Nearly the same trends are found for the \textit{z}-band GCLF. 
We perform linear and exponential least-squares fits to the radial trends of $M_{g,\rm TO}$ and $\sigma_g$, respectively.
The best-fit relations are indicated as red dashed curves in \reffig{GCLF_R}:
\begin{equation}
    M_{{g,\rm TO}} = ( 0.003\pm0.001)\times R+(-7.319\pm0.020)   
\end{equation}
\begin{equation}
    \sigma_{{g}} = (1.05 \pm 0.08) + (0.33 \pm 0.08) \exp\left(-\dfrac{R}{19.3 \pm 15.6}\right)
\end{equation}
where $R$ is in unit of arcminutes.
As illustrated in the left panel of \reffig{GCLF_R}, in the bright half of the GCLF, the decrease in $\sigma_{g}$ is accompanied by a declining fraction of GCs at the brighter (more massive) end of the distribution toward larger galactocentric radii. Thus both the increase in $M_{g,\rm TO}$ and the decrease in $\sigma_g$ suggest mass stratification of the GCS.

\subsection{Color gradient}\label{colorgradient}

\begin{figure*}
    \centering
    \includegraphics[width=0.99\linewidth]{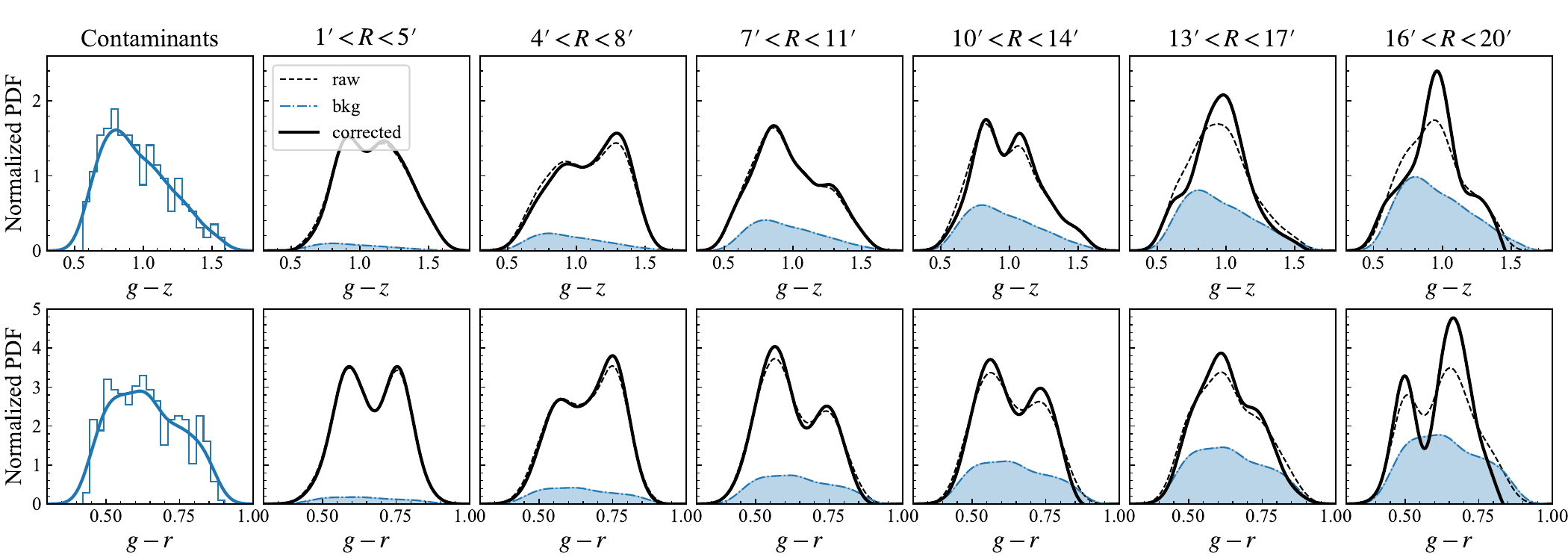}
    \caption{GC color distributions.
    The leftmost column shows the normalized color distributions of contaminants, with blue solid curves constructed using Gaussian KDE. 
    The contaminant sample is defined as candidates located between $40^\prime$ and $70^\prime$.
    The remaining columns show the normalized raw (dashed) and background-corrected (solid) color distributions in six radial bins, with radial ranges indicated in the panel titles.
    The blue dash-dotted curves and shaded regions represent the expected background distribution scaled by the corresponding contaminant fractions, which are subtracted from the raw distributions to obtain the background-corrected ones. }
    \label{Color_correction}
\end{figure*}
\begin{figure*}
    \centering
    \includegraphics[width=0.95\linewidth]{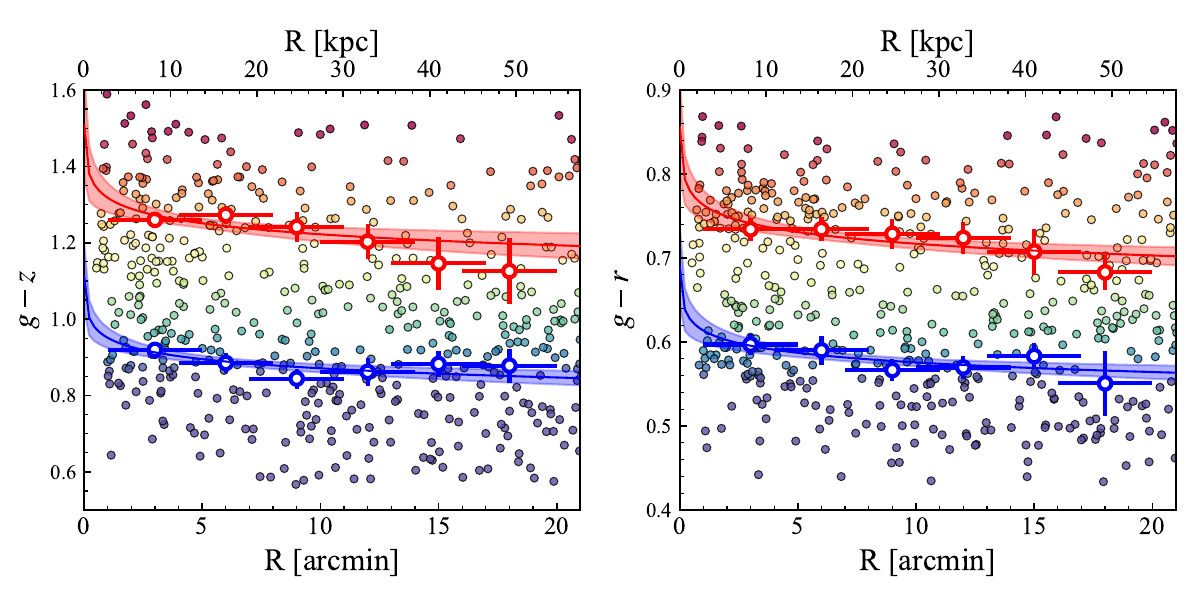}
    \caption{The radial color variations of GC candidates.
    The $g-z$ and $g-r$ colors are shown in the left and right panels, respectively.
    The little dots are individual GC candidates, plotted according to the $g-z$ and $g-r$ colors.
    The large dots represent contaminant-corrected peak colors of the GC distributions at different radii, with horizontal error bars representing the radial range and vertical error bars obtained from bootstrapping for 100 times.
    The best-fit linear dependence of peak colors on $\log R$ are shown as solid curves. See Section \ref{colorgradient} for details.}
    \label{Color_grad}
\end{figure*}

Broadband colors in the optical or near-IR wavelengths primarily reflect the metallicity of GCs \citep[e.g.,][]{Brodie_2006}. Accretion of satellite galaxies may deposit proportionally more metal-poor (blue) GCs at large galactocentric radii of the host galaxies \citep{Cote1998}, which may result in the spatial distribution difference between GCs of different metallicities or colors \citep[e.g.,][]{Faifer_2011,Jennings_2014,Cantiello2015}.
Thus the radial color gradient of GCS serves as an important indicator of assembly history of the host galaxy.
Here we fit $g-z$ and $g-r$ color distributions of GCs at various radii and determine their peak locations.

We divide our candidates into radial bins, centered at $3^\prime$, $6^\prime$, $9^\prime$, $12^\prime$, $15^\prime$, and $18^\prime$, each with a width of $4^\prime$.
The raw color distributions of GC candidates in these bins are constructed using Gaussian KDE (implemented in \textsc{Scikit-learn} \citep{scikit_learn}), with kernel size of 0.08 mag for $g-z$ and 0.04 mag for $g-r$, shown as dashed curves in \reffig{Color_correction}.
Since the contaminants might affect the color distributions of genuine GCs, we correct the raw color distributions through the following steps.
First, a contaminant sample is defined as all candidates within the annulus between $40^\prime$ and $70^\prime$, where the contamination fraction exceeds 96\%. This fraction is calculated based on the number of genuine GCs predicted by the best-fit Sersic profile within the same radial range. The color distribution of the contaminant sample is constructed using KDE and shown in the leftmost panels of \reffig{Color_correction}. Second, the raw color distribution of GC candidates in each radial bin is then corrected by subtracting a scaled contaminant distribution (blue dash-dotted curve and shading in the corresponding panel). The scaling for each radial bin is based on its contamination fraction. The corrected distributions are plotted as black solid curves in \reffig{Color_correction}. Note that all distributions shown in \reffig{Color_correction} are normalized except for the scaled contaminant distribution in each radial bin.

After the contaminant correction, we employ Gaussian Mixture Model (GMM) to determine the blue and red peaks of the corrected color distributions, implemented with \textsc{Scikit-learn}, with the number of components fixed to two.
Applying the GMM to the fiducial GCs yields a blue peak at $g-z = 0.91$, a red peak at $g-z = 1.26$, and a division threshold of $g-z = 1.08$ for the two sub-samples.
For each radial bin, we first draw a set of random samples from the corrected color distribution, ensuring its sample size matches the number of expected bona fide GCs in this bin.
Then we use GMM with two components to model the distribution of the random sample and obtain the red and blue peak locations.
The values and errors of the peak locations are obtained by repeating the random sampling and GMM modeling process for 100 times. While we do not attempt to assess the significance of the color bimodality, we note that it gradually diminishes at $R$ $\gtrsim$ 15$^{\prime}$, as illustrated in \reffig{Color_correction} and indicated by the large uncertainties of the two color peaks in \reffig{Color_grad}.

It is clear that both $g-z$ and $g-r$ peak colors become bluer at larger radii for both blue and red GCs, as shown in \reffig{Color_grad}.
Colors of individual GC candidates are plotted against their galactocentric radii, while the most probable peak colors are represented by the red-edged and blue-edged squares.
The least-squares fitting to the peak locations gives the color gradients, which are indicated by the solid curves: 
\begin{equation}
    (g-z)_{{\rm blue}} = ( -0.08\pm0.03)\times \log R+(0.99\pm0.04) 
\end{equation}
\begin{equation}
    (g-z)_{{\rm red}} = ( -0.10\pm0.05)\times \log R+(1.36\pm0.06) 
\end{equation}
\begin{equation}
    (g-r)_{{\rm blue}} = ( -0.04\pm0.02)\times \log R+(0.63\pm0.02) 
\end{equation}
\begin{equation}
    (g-r)_{{\rm red}} = ( -0.05\pm0.02)\times \log R+(0.78\pm0.03) 
\end{equation}
where $R$ is the galactocentric radius in unit of kpc.
The $g-z$ gradients are larger than but consistent within uncertainty with those given by \cite{Jennings_2014} based on GC candidates within a central radius of 14 kpc in NGC 3115: $(g-z)_{\rm blue}\propto(-0.06\pm 0.02)\times\log R$ and $(g-z)_{\rm red}\propto(-0.05\pm0.04)\times\log R$.

\subsection{Substructures of the GCS}\label{substructures}

\begin{figure*}
    \centering
    \includegraphics[width=0.95\linewidth]{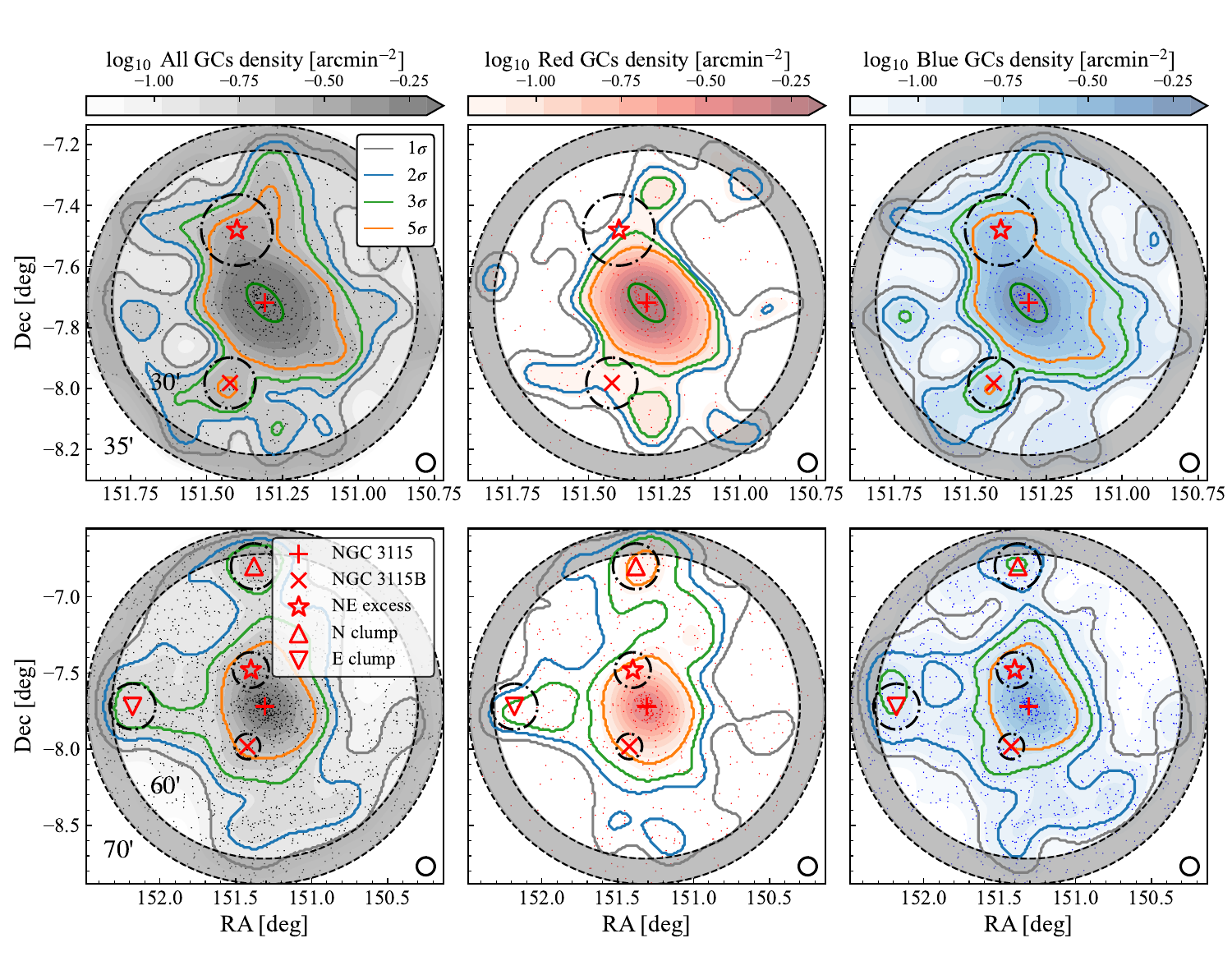}
    \caption{Projected density maps of GC candidates showing substructures.
    The maps are shown at two different scales: $35^\prime$ (top) and $70^\prime$ (bottom), and for three GC subsamples: all candidates (left), red candidates (middle), and blue candidates (right). 
    These distributions are measured using Gaussian kernel density estimation (KDE), with the kernel sizes indicated by the small black circles in the lower-right corners of each panel.
    The contours represent 1$\sigma$ (gray), 2$\sigma$ (blue), 3$\sigma$ (green), and $5\sigma$ (orange) above the mean background density, which is estimated within the shaded annular regions.
    The green ellipses in the top panels represent the isophote of $\mu_g=25\ {\rm mag}/{\rm arcsec}^2$ measured on \textit{g} band image.
    The center of NGC 3115 and 3115B are indicated as a red plus symbol and a red cross, respectively.
    The NE excess, N clump, and E clump are marked by the red star, a red triangle, and a red inverted triangle, respectively.}
    \label{substruc}
\end{figure*}

A more straightforward way to test the hierarchical assembly process is to search for tidal debris or sub-structures resulting from past or ongoing accretion events. The relatively high purity and full spatial coverage of our GC candidate sample enable the effective detection of substructures in the GCS of NGC 3115.

We identify sub-structures of GCS in the density maps of GC candidates at two spatial scales: the central $35^\prime$ region and the full $70^\prime$ field, shown in top and bottom panels of \reffig{substruc}, respectively.
The middle and right panels display only red and blue candidates, respectively.
The red and blue sub-samples are divided by the threshold $g-z=1.08$, which is determined by fitting the color distribution of fiducial GCs using GMM.
The density maps are constructed using Gaussian KDE with bandwidths of $3.5^\prime$ and $7^\prime$ for the $35^\prime$ and $70^\prime$ scales, respectively.
To identify substructures, we first estimate the background densities within outer annuli between $30^\prime\sim35^\prime$ and $60^\prime\sim70^\prime$, respectively, for top and bottom panels. Then contours of $1\sigma$, $2\sigma$, $3\sigma$, and $5\sigma$ above the mean background density are highlighted in \reffig{substruc}.
The overdensities with at least $3\sigma$ significance are considered as detections of sub-structures.

Within the inner $35^\prime$ region, a clumpy structure at the southeast is noticeable with a peak significance of $5\sigma$.
It is associated with the dwarf galaxy NGC 3115B (or NGC 3115 DW1), whose GCS has been extensively explored \citep{Hanes_1986,Durrell_1996,Puzia_2000,Kundu_2001}.
Its photometric center is marked by the red diagonal cross.
More interestingly, we identify an asymmetry in the main body of the GCS, with an excess toward the northeast at a significance over $5\sigma$ (NE excess, red star), and much fewer GC candidates are found in the opposite side of NGC 3115.
The NE excess mainly lies along the major axis of NGC 3115, but it exhibits a tendency to shift northward.

On the larger $70^\prime$ scale, the NE excess and the NGC 3115B clump become smoothed out due to the larger smooth kernel.
However, two new overdensities emerge farther to the north (N clump, red triangle) and east (E clump, red inverted triangle), each with a significance exceeding $3\sigma$.

Basic information of these sub-structures is given in Table \ref{tab:struc}.
We count the number of candidates in each sub-structure within a radius of $R^*$ around its center position.
Here, the radii are $R^*=5^\prime$, $7^\prime$, $9^\prime$, and $9^\prime$ for the four substructures, respectively. They are visually determined to roughly enclose the $3\sigma$ areas and they are indicated by the dash-dotted circles in \reffig{substruc}.
The estimates of contaminant numbers are also listed as $N_{\rm bkg}$, which are calculated by multiplying the mean background density and the areas of the circular regions of corresponding sub-structures.

\begin{table}[h]
\centering
\caption{Substructures. } \label{tab:struc}
\begin{threeparttable}
\begin{tabularx}{0.5\textwidth}{lccccccc}
\toprule
Name & RA & Dec & $R^*$\tnote{a} & $N$ & $N_{\rm red}$ & $N_{\rm blue}$ & $N_{\rm bkg}$ \\
\midrule
NGC 3115B & 151.42 & -7.98 & $5^\prime$ & 23 & 6  & 17 & 8    \\
NE excess & 151.40 & -7.48 & $7^\prime$ & 50 & 9  & 41 & 16    \\
N clump   & 151.38 & -6.80 & $9^\prime$ & 61 & 23 & 38 & 26  \\
E clump   & 151.18 & -7.72 & $9^\prime$ & 48 & 16 & 32 & 26  \\
\bottomrule
\end{tabularx}
\begin{tablenotes}
    \footnotesize
    \item[a] The radius used to count the number of candidates.
\end{tablenotes}
\end{threeparttable}
\end{table}

The discovery of NE excess, N clump, and E clump suggests an ongoing assembly of the stellar halo of NGC 3115. Given the relative isolation of the NGC 3115 group, these sub-structures likely result from recent satellite accretion events.
However, no clear counterpart in diffuse stellar light is detected in the Legacy Survey images. Deeper imaging in the future may help reveal the diffuse stellar component associated with these accretion events. We mention that, debris from disrupted satellite galaxies (if any) may become unrecognizable in configuration space after a few orbital periods due to phase mixing \citep{Helmi1999}. 
A relatively comprehensive census of satellite accretion events, along with a thorough understanding of the sub-structures identified here, requires velocity and/or chemical information of the GCs.

\section{Discussion} \label{sec:dis}

\begin{figure}
    \centering
    \includegraphics[width=0.99\linewidth]{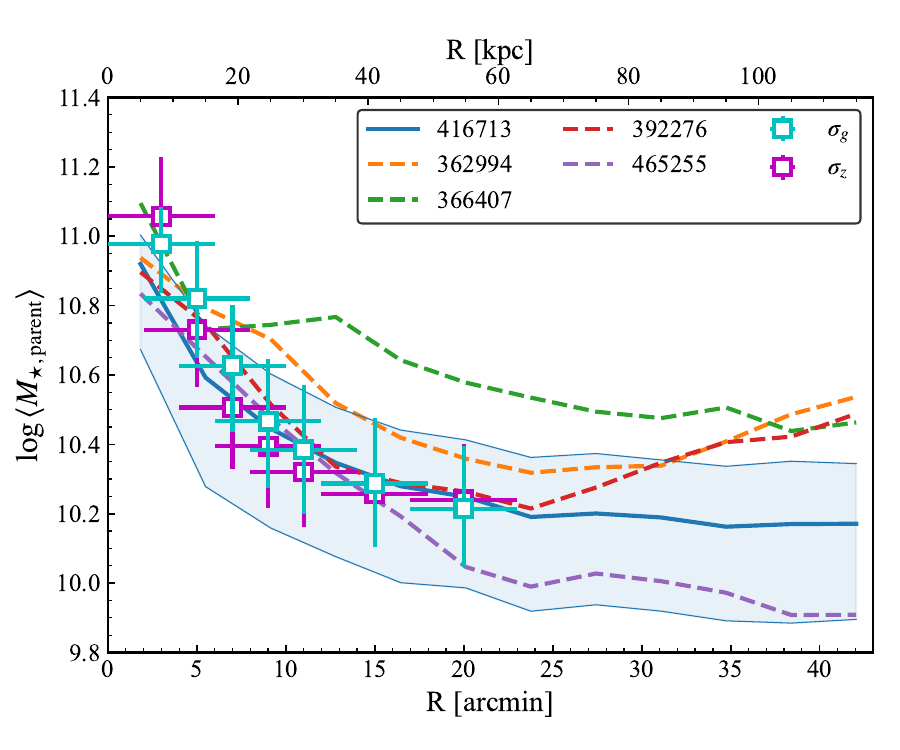}
    \caption{Radial variations of the average stellar mass of parent galaxies. The cyan-edged and magenta-edged squares represent the galaxy masses estimated from the dispersions of GCLFs in the \textit{g} and \textit{z} bands, respectively (see Section \ref{gclf2growthhistory} for details). The solid or dashed curves represent the weighted average for five TNG50 subhalos that match the basic properties of NGC 3115. The blue-shaded region represents the 68\% confidence interval for the subhalo plotted as solid curve (see Section \ref{decode} for details).}
    \label{galMass}
\end{figure}

\subsection{The assembly history traced by GCS mass stratification} \label{gclf2growthhistory}

As demonstrated above, we have observed evident mass stratification in the GCS of NGC 3115, through both the increasing effective radius with decreasing \textit{g} mag (Section \ref{radial}) and the narrowing GCLF with increasing galactocentric radius (Section \ref{gclf_radial}).

Mass stratification can arise from multiple mechanisms, depending on the nature of the system.
In stellar clusters, two-body relaxation causes more massive stars to sink inward and less massive stars migrate outward \citep{McMillan_2007}.
In galaxy clusters, dynamical friction plays a key role in driving more massive galaxies to infall faster than low-mass ones \citep{Contini_2015,Bosch_2016}.
As for GCs in galaxies, dynamical friction-driven orbital decay is efficient only for the massive GCs in dwarf galaxies or in the very central regions of massive galaxies \citep{Oh_2000,Goerdt_2006,Bekki_2010,Gabriella_2024,Lyu2025}. 
The timescale for orbital decay of GCs due to dynamical friction can be estimated using the classical expression \citep[see Equation 8.13 of][]{Binney_2008}:
\begin{equation}
    t_{\mathrm{fric}} \approx \frac{1.17}{\ln \Lambda} \frac{r^2 v_c}{G M}
\end{equation}
By assuming an isothermal dark matter halo, adopting a rotation velocity of $v_c\simeq200$ km s$^{-1}$ for NGC 3115 \citep{Guerou_2016}, and a typical value of Coulomb logarithm $\ln \Lambda\sim10$, the dynamical friction timescale of a $10^{6}\ \rm M_{\odot}$ cluster reaches beyond a Hubble time at galactocentric radii $>$ 1.5 kpc. 
This timescale is inversely proportional to the mass of a GC and directly proportional to the square of its galactocentric radius. 
Therefore, the mass stratification of NGC 3115, observed out to at least $\sim60$ kpc (\reffig{GCLF_R}), cannot be attributed to dynamical friction. Besides dynamical friction, tidal evaporation may also affect the GC mass distribution, and it is more effective for lower-mass GCs. It alters both the turnover magnitude and dispersion of the GCLF by gradually eroding the low-mass end through two-body relaxation in the galactic tidal field \citep{Baumgardt2003}.
The stronger tidal evaporation effect at smaller galactocentric distances may lead to a brighter GCLF turnover magnitude and a narrower dispersion, the latter of which contradicts observations of NGC 3115. Moreover, extensive simulations \citep{Baumgardt2003, Gieles2023} suggest that tidal evaporation has minimal impact on the higher-mass half of the GCLF, where mass stratification of NGC 3115 is observed.

Therefore, the observed GCS mass stratification most likely serves as an indicator of the growth history of NGC 3115, rather than as a result of the internal evolution of the GCS itself.
The turnover magnitude and especially the dispersion of the GCLF are known to correlate with the luminosity and mass of the host galaxy \citep{Jordan_2007, Villegas2010}, with lower-mass galaxies exhibiting fainter turnover magnitudes and smaller dispersions. Our findings thus suggest that the dominant contributors to the stellar halo decrease in mass with increasing galactocentric distance from NGC 3115, reflecting an inside-out growth driven by the accretion of, on average, progressively lower-mass satellites at larger radii.

To be quantitative, we estimate the average stellar mass of the parent galaxies of GCs at different galactocentric radii, by utilizing the relation between GCLF dispersion (in both \textit{g} band and \textit{z} band) and galaxy {\em B}-band luminosity obtained by \cite{Jordan_2007} and adopting a typical {\em B}-band mass-to-light ratio of 2. The results are shown as cyan-edged squares for the \textit{g} band and magenta-edged squares for the \textit{z} band in \reffig{galMass}. The two bands yield consistent results within the uncertainties. Note that we use the dispersion trend instead of the turnover magnitude trend, as it exhibits a much stronger and steeper correlation with galaxy luminosity. 

As shown in \reffig{galMass}, the average parent galaxy mass decreases from $\sim10^{11}\rm\ M_\odot$ within 10 kpc to $\sim10^{10.2}\rm\ M_\odot$ beyond 20 kpc. We note that the GCLF, and consequently the inferred average galaxy mass at a given radius, reflects contributions from both the primary progenitor (formed during the early dissipative phase) and accreted satellites, effectively representing a GC-weighted average. 
The global GCLF of a galaxy is largely shaped by its inner galactocentric regions, as is the case for NGC 3115. The average parent galaxy mass of the central 10 kpc approaches that of NGC 3115 as a whole, suggesting that the GCs in this region predominantly originate from the primary progenitor of NGC 3115, which presumably formed through violent dissipative process during the early epochs, preceding the ``second phase'' of gradual accretion of satellite galaxies.

Besides GCS mass stratification, the inside-out two-phase formation history may also naturally explain the notable bimodal GC metallicity distribution of NGC 3115 \citep{Brodie_2012}, where the (red) metal-rich GCs formed mostly in the first violent dissipative phase while the (blue) metal-poor GCs were mostly accreted from satellite galaxies \citep{Ashman1992,Cote1998,Tonini_2013}. 
While the classical two-phase scenario associates red GCs primarily with in-situ formation and blue GCs with ex-situ accretion, more recent studies suggest that both subpopulations may contain a mix of in-situ and accreted origins \citep[e.g.,][]{Forbes_2018,Kruijssen2019,Keller2020}. Moreover, observational studies have shown that, toward the lower-luminosity end of galaxies, the colors of both the blue and red peaks (when present) in the GC color distribution shift to bluer values (albeit with substantial scatter), while the red peak gradually diminishes \citep[e.g.,][]{Peng2006,Choksi2019}. Therefore, the progressively lower average parent-galaxy mass at larger galactocentric radii naturally accounts for the weak negative radial color gradients observed in both the blue and red subpopulations of NGC 3115 (Section \ref{colorgradient}).

\subsection{Decoding the assembly history with simulations}\label{decode}

\begin{figure*}
    \centering
    \includegraphics[width=0.95\linewidth]{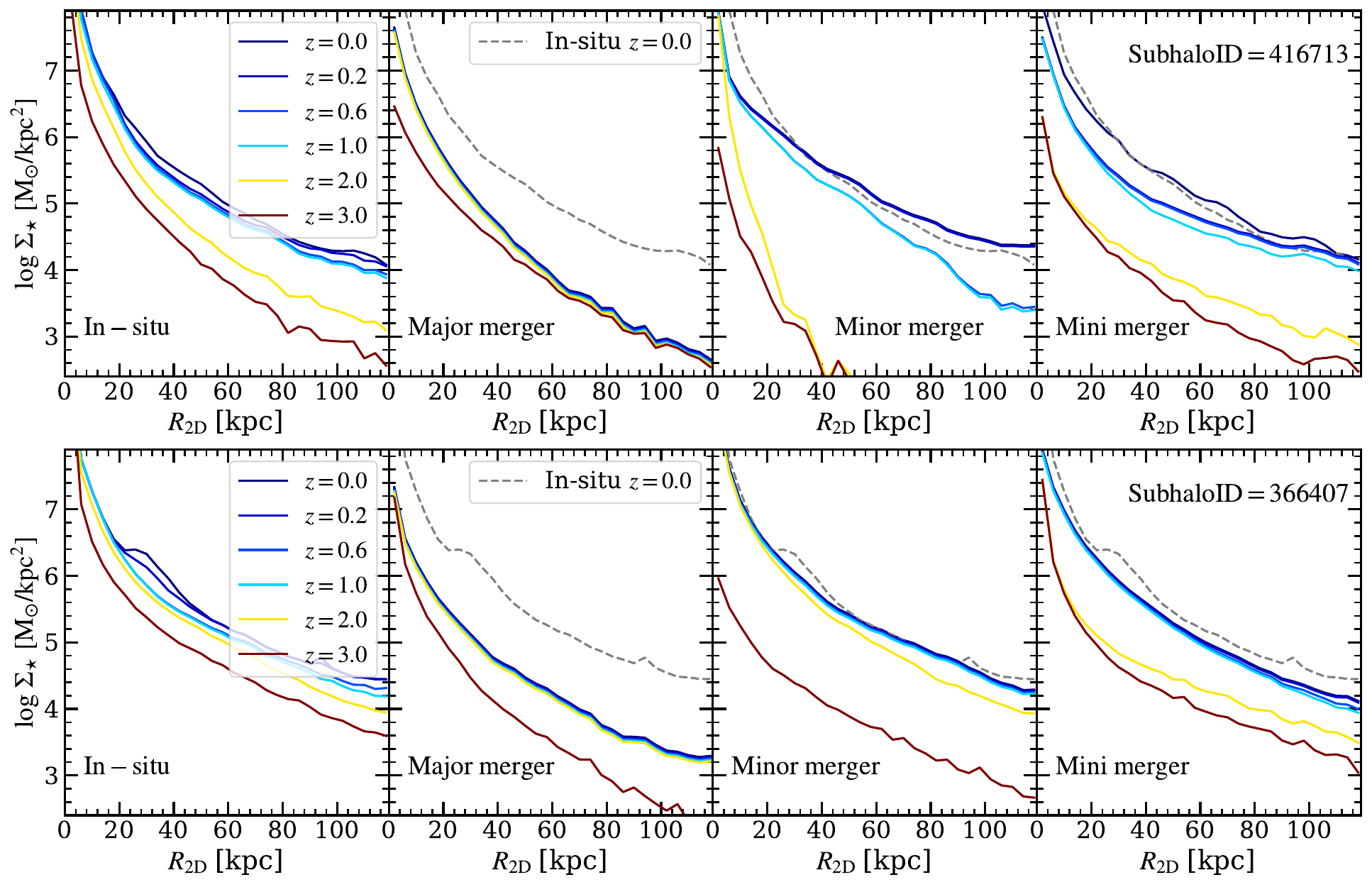}
    \caption{The assembly histories of the TNG50 subhalos that most closely (top panels) and least closely (bottom panels) match the radial profiles of the average parent galaxy mass of NGC 3115 shown in Figure \ref{galMass}.
    The first column presents the evolution of the stellar mass surface density profiles for stars formed \textit{in situ}. The second, third, and fourth columns show the profiles for stars accreted through major mergers (secondary-to-primary stellar mass ratio $\mu > 1/4$), minor mergers ($1/10 < \mu < 1/4$), and mini mergers ($\mu < 1/10$), respectively. See Section \ref{decode} for more details.}
    \label{assemblyhis}
\end{figure*}  

To decode the assembly history of NGC 3115 based on the average parent galaxy stellar mass of GCs as a function of galactocentric radius, we utilize the TNG50 to select NGC 3115-like subhalos in simulations, as described in Section \ref{tng50}.
Estimation of the GC-weighted average stellar mass of progenitor parent galaxies $\langle M_{\star, {\rm parent}}\rangle$ is described in Section \ref{sec:avmass_tng50} and the results are overplotted in \reffig{galMass}. Note that the weights used for calculating the average correspond to the fraction of GCs contributed by each parent galaxy at a given radius. We emphasize that the simulations here do not precisely replicate the average parent mass inferred from the GCLF of NGC 3115. However, it is sufficient for an exploratory investigation prior to the development of a sophisticated modeling of massive galaxy GCS assembly. Our aim here is to provide qualitative insights rather than an exact reproduction of the observational results. 

The $\langle M_{\star, {\rm parent}}\rangle$ profiles of the five TNG50 subhalos are overplotted in \reffig{galMass}. 
Among them, the profile of subhalo-416713, which most closely resembles that of NGC 3115 based on visual inspection, is shown with a thick solid curve for the median trend and a blue-shaded region representing the 68\% confidence interval based on multiple probabilistic sampling of the GC specific frequency distributions (Section \ref{sec:gctag_tng50}). The median profiles for the other four subhalos are depicted with dashed curves. It is noteworthy that three of the five subhalos have radial profiles in agreement with the observed profiles within uncertainties, in line with the theoretical expectation of a broad correlation of merger histories with bulge fraction and galaxy mass (as the primary criteria for selecting NGC 3115-like subhalos), albeit with substantial scatter \citep{Martig2012,Rodriguez-Gomez2016}.

To demonstrate the constraining power of $\langle M_{\star, {\rm parent}}\rangle$ profiles, the assembly histories of the best-matching subhalo and the least-matching subhalo are shown in the upper and lower panels of \reffig{assemblyhis}, respectively. For the best-matching subhalo, the radial stellar surface density profiles resulting from either {\em in situ} formation or {\em ex situ} accretion flatten over cosmic time, reflecting an inside-out growth. In contrast, the least-matching subhalo exhibits nearly self-similar growth across radii for both {\em in situ} and {\em ex situ} contributions. Moreover, the stellar contribution from minor or mini mergers dominates over the {\em in situ} component across nearly the entire system of the best-matching subhalo, whereas the opposite is true for the least-matching subhalo.


\section{Summary} \label{sec:sum}

In this work, we utilize the wide-field \textit{g}, \textit{r}, \textit{z} images from Legacy Surveys to identify GC candidates across a $70^\prime$ ($\sim190$ kpc; $\sim$ 1.1$R_{\rm vir}$) radius around NGC 3115. Gaia is also employed as a useful auxiliary. From these data, we derive robust radial GC number-density profiles out to $30^\prime$ ($\sim$ 0.5$R_{\rm vir}$), and identify significant substructures of the GC system out to $60^\prime$. Our sample comprises 731 GC candidates within the central $30^\prime$, achieving a purity of $60\pm4\%$ and a completeness exceeding 92.5\% at $g=23.5$ mag ($M_{g}=-6.4$ mag).

Leveraging this highly complete dataset over an unprecedented sky coverage, we report the discovery of mass stratification in the GCS of NGC 3115.
On the one hand, the effective radius of GCS increases with decreasing \textit{g} mag, suggesting more extended distributions of fainter GCs.
On the other hand, the turnover magnitude and dispersion of GCLF increases and decreases, respectively, with increasing galactocentric radius, indicating proportionally less bright GCs at larger radii.
These results represent the first observational evidence of GCS mass stratification in a massive galaxy.
In addition, our analysis reveals overall negative color gradients for both the blue and red GCS, extending out to a galactocentric distance of 50 kpc. We also identify several GC substructures extending out to a radial distance of 160 kpc, suggesting an ongoing assembly of the stellar halo of NGC 3115.

The GCS mass stratification revealed in this study is interpreted as a powerful probe of the two-phase assembly scenario for massive galaxies—characterized by an early, rapid {\em in situ} formation followed by the prolonged accretion of satellite galaxies. 
We convert the observed radial variation of the GCLF in NGC 3115 into a radial profile of the GC-weighted average parent galaxy mass, using the empirically established correlation between GCLF width and host galaxy luminosity. The resulting radial trend in average parent mass for NGC 3115 shows encouraging consistency with that of TNG50 subhalos of similar mass and bulge-to-total mass ratios. These NGC 3115-like subhalos in TNG50 are indeed commonly characterized by an inside-out assembly history.

Ground-based imaging studies of extragalactic GCSs are generally hindered by significant contamination from foreground Milky Way stars and background galaxies. This limits robust analysis to either very nearby galaxies, where foreground contamination can be largely mitigated, or to small galactocentric radii ($\lesssim$ 60 kpc in this study), where the number density of GCs remains higher than, or at least comparable to, the level of contamination. 
However, the advent of space-based wide-field telescopes, such as Chinese Space Station Telescope (CSST) and {\em Euclid}, will enable analyses like the one presented here with greatly improved sample purity and completeness for a large sample of nearby galaxies. This will provide comprehensive constraints on the assembly histories of massive galaxies in general.

\begin{acknowledgments}
\small
HRD appreciates Zehao Zhang for constructive discussion in the candidates selection.
This work has been supported by the National Key Research and Development Program of China (No. 2023YFC2206704) and the China Manned Space Program with grant No. CMS-CSST-2025-A04. 
We also acknowledge support from the National Key Research and Development Program of China (grant No. 2023YFA1608100), the NSFC (grant Nos. 12122303, 11973039 and 12192224), the China Manned Space Project (grant Nos. CMS-CSST-2021-B02 and CMS-CSST-2021-A07), the CAS Pioneer Hundred Talents Program, the Strategic Priority Research Program of Chinese Academy of Sciences (grant No. XDB 41000000), and the Cyrus Chun Ying Tang Foundations.
HYW acknowledgements supports from CAS Project for Young Scientists in Basic Research, Grant No. YSBR-062. 

This work has made use of data from the European Space Agency (ESA) mission {\it Gaia} (\url{https://www.cosmos.esa.int/gaia}), processed by the {\it Gaia} Data Processing and Analysis Consortium (DPAC, \url{https://www.cosmos.esa.int/web/gaia/dpac/consortium}). Funding for the DPAC has been provided by national institutions, in particular the institutions participating in the {\it Gaia} Multilateral Agreement.

This research has made use of data from the Legacy Surveys.The Legacy Surveys consist of three individual and complementary projects: the Dark Energy Camera Legacy Survey (DECaLS; Proposal ID \#2014B-0404; PIs: David Schlegel and Arjun Dey), the Beijing-Arizona Sky Survey (BASS; NOAO Prop. ID \#2015A-0801; PIs: Zhou Xu and Xiaohui Fan), and the Mayall z-band Legacy Survey (MzLS; Prop. ID \#2016A-0453; PI: Arjun Dey). DECaLS, BASS and MzLS together include data obtained, respectively, at the Blanco telescope, Cerro Tololo Inter-American Observatory, NSF’s NOIRLab; the Bok telescope, Steward Observatory, University of Arizona; and the Mayall telescope, Kitt Peak National Observatory, NOIRLab. Pipeline processing and analyses of the data were supported by NOIRLab and the Lawrence Berkeley National Laboratory (LBNL). The Legacy Surveys project is honored to be permitted to conduct astronomical research on Iolkam Du’ag (Kitt Peak), a mountain with particular significance to the Tohono O’odham Nation.
\end{acknowledgments}


\bibliography{ms}{}
\bibliographystyle{aasjournalv7}

\end{CJK*}
\end{document}